\documentclass[aps,twocolumn,pra,superscriptaddress,showpacs,tightenlines]{revtex4}
\usepackage{epsfig,graphicx,times}
\usepackage{amstext}
\usepackage{amsmath}
\usepackage{amssymb}
\usepackage{graphicx}
\usepackage{latexsym}
\usepackage{bm}

\providecommand{\U}[1]{\protect\rule{.1in}{.1in}}

\begin{document}

\title{Multi-Atomic Mirror for Perfect Reflection of Single Photons in A Wide Band of Frequency}

\author{Yue Chang}
\affiliation{Institute of Theoretical Physics, Chinese Academy of
Sciences, Beijing, 100190, China}

\author{Z. R. Gong}
\email{gongzr@itp.ac.cn}
\affiliation{Institute of Theoretical
Physics, Chinese Academy of Sciences, Beijing, 100190, China}

\author{C. P. Sun}
\email{suncp@itp.ac.cn} \homepage{http://www.itp.ac.cn/~suncp}%
\affiliation{Institute of Theoretical Physics, Chinese Academy of
Sciences, Beijing, 100190, China}

\begin{abstract}
A resonant two level atom doped in a one-dimensional waveguide
behaves as a mirror, but this  single-atom ``mirror" can only
reflect single photon perfectly at a specific frequency. For a one
dimensional coupled-resonator waveguide, we propose to extend the
perfect reflection region from a specific frequency point to a wide
band by placing many atoms individually in the resonators in a
finite coordinate region of the waveguide. Such a doped resonator
array promises to control the propagation of a practical photon wave
packet with certain momentum distribution instead of a single
photon, which is ideally represented by a plane wave with specific
momentum. The studies based on the discrete-coordinate scattering
theory display that such hybrid structure with finite atoms indeed
provides a near-perfect reflection for a single photon in a wide
band. We also calculated the photon group velocity distribution,
which shows that the perfect reflection wide band exactly
corresponds to the stopping light region.
\end{abstract}

 \maketitle

\section{Introduction}

Quantum manipulation in all-optical fashions \cite%
{lukin,liyong1,liyong2,gong} is very crucial to the future development of
high technology concerning optical communication \cite%
{lukin,wall,fan,zhou,zhou3,zhou4,liao}, quantum information process \cite%
{lukin,wall,zhou,weber,infor}, and the next-generation quantum devices,
e.g., single-photon transistors \cite{wall,weber,orrit}, quantum switches
\cite{lukin,fan,zhou,zhou3,zhou4,liao,orrit,zhou5,liao1}, and photon
storages \cite{liyong1,liyong2,gong,zhou1,zhou2}.

The core physics behind all-optical quantum manipulation is to explore the
single-photon scattering and propagation in the confined structure of the
sizes comparable to the wavelength of the photon. The investigations about
the single-photon propagation in one-dimension involve some new phenomena,
such as the perfect reflection of the single photon by atomic mirror \cite%
{lukin,fan,zhou,zhou5,olshanii}, slowing light processes \cite%
{liyong1,liyong2,zhou1,zhou2,hughes1} in hybrid coupled-resonator
waveguides, and other related issues \cite{gong,wall,fan,zhou4,law}. The
hybrid structures concerned here could be implemented physically with linear
defect cavities in photonic crystals \cite{green}\ with doped quantum dots,
or superconducting transmission line resonators \cite{zhou3,liao,liao1}
coupled to a superconducting qubit \cite{wall,you1,you2,chio,you}. These
physical systems with artificial band structures coupled to a two-level
system enable us to control the transport of a single photon. By tuning the
structure parameters of the hybrid system, the two-level system acts as a
quantum switch, making the transporting single photon be reflected
perfectly, or transmit totally. In this sense, the two-level system can
behave as an ideal mirror \cite{lukin,fan,zhou,zhou5,olshanii}.

It has been shown that a single-photon transistor using nanoscale surface
plasmons \cite{lukin} coupled with a three level atom in EIT
(electromagnetically induced transparency) setup \cite%
{liyong1,liyong2,gong,zhou1,zhou2}, exhibits controllable behavior in the
transmission spectra. We have re-examined the coherent transport of a single
photon in a coupled-resonator array coupled to a controllable two-level
system \cite{zhou}. Being different from the linear dispersion relation in
Chang \textit{et al.}'s setups \cite{lukin,fan}, the cosine-type dispersion
\cite{zhou} will result in two bound-states in the hybrid system. The total
reflection by the two-level system, which behaves as an ideal mirror, has
been found associated with the Fano-Feshbach and Breit-Wigner line shapes
\cite{landau} around the resonance in the reflection spectrum. However, in
all these works \cite{lukin,fan,zhou}, we emphasize that the perfect
reflection exists only at a specific frequency point. It brings physical
difficulties in practical applications, such as the efficiency to control an
optical pulse, which actually is a superposition of the plane waves with
different frequencies where the off-resonant components could deviate from
the perfect reflection point dramatically.

In this paper, we propose an experimentally accessible setup based on the
coupled-resonator array with doped atoms hybrid system, which is excepted to
realizes perfect reflection with wide spectrum, and thus can perfectly
reflect an optical pulse, namely, a single photon wave packet. Here, we use
a \textquotedblleft thick\textquotedblright\ atomic mirror which is made of
an array of two-level atoms individually doped in some cavities arranged in
a coordinate region of the one-dimensional coupled-cavity waveguide. The
physical mechanism is intuitive: when a photon little far from resonance
reaches one cavity coupled to the doped atom, it is reflected by the atom
partly, and then the left part passing through the next atom experiences the
same process. This process is repeated many times, which may realizes the
perfect reflection with wide spectrum so long as the interference
enhancement could be suppressed by some mechanism. The emergence of
wide-band spectrum is also shown schematically in Fig. 1, where the perfect
reflection region is from a specific incident energy point (for single atom)
to a wide band (for more than one atoms).

In details, we will study the wide-band scattering phenomena for our
proposed atomic mirror by using the discrete coordinate scattering approach
\cite{zhou,zhou5}. Here, with the second order processes for the atom
absorbing a photon and then radiating back inside cavity, the basic role of
the doped atoms is to provide an effective potential like a local resonant
Dirac comb \cite{flugge}, which leads to the stopping and slowing light
phenomena \cite{liyong1,liyong2,zhou1,zhou2,hughes1}. By detuning the
coupling strength of the atom coupled to the single cavity mode, we can
feasibly control the width of the perfect reflection band. It is noticed
that this wide-band spectrum phenomena have been implied in some works \cite%
{zhou5,hughes,law}.

The paper is organized as follows. In Sec. II, we propose our model and
solve it with the discrete coordinate scattering theory. In Sec. III, we
study the microscopic physical mechanism and acquire the slowing light
phenomenon in Sec. IV. In Sec. V, we study the influence of the
imperfections in experiments on the perfect reflection wide-band. At last,
we give a summary in Sec. VI.

\section{Wide-Band Atomic Mirror for Single Photon in One Dimension}

\begin{figure}[ptb]
\includegraphics[bb=32 446 527 645, width=8 cm, clip]{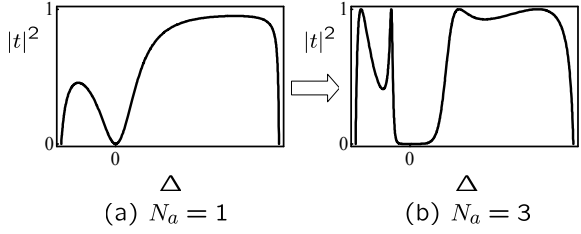}
\caption{Schematic diagram for the emergence of the wide band for perfect
reflection. Here, $\Delta$ is the detuning concerning the incident photon's
energy, and $\left\vert t\right\vert ^{2}$ is the transmission coefficient.
It is shown that for single atom [for (a)], the perfect reflection region is
only a specific point; while when the atom number $N_{a}$ is 3, the region
is extended to a wide band.}
\end{figure}

Our hybrid system in this paper is shown in Fig. 2, where the $N_{a}$
two-level atoms are individually embedded in a one-dimensional
coupled-resonator waveguide (CRW) \cite{nori} . The atoms play essential
role in controlling the propagation of a single photon. The Hamiltonian $%
H=H_{c}+H_{I}$ of this hybrid system\ consists of two parts, the CRW part
described by a tight-binding boson model%
\begin{equation}
H_{c}=\omega \sum_{j=-N}^{N}a_{j}^{\dag }a_{j}+V\sum_{j=-N}^{N}\left(
a_{j}^{\dag }a_{j+1}+a_{j+1}^{\dag }a_{j}\right) ,  \label{2}
\end{equation}%
and the part of two-level atoms interacting with the cavity fields%
\begin{equation}
H_{I}=\frac{\Omega }{2}\sum_{j=1}^{N_{a}}\left( \sigma _{j}^{z}+1\right)
+g\sum_{j=1}^{N_{a}}\left( a_{j}\sigma _{j}^{+}+a_{j}^{\dag }\sigma
_{j}^{-}\right) .  \label{3}
\end{equation}

\begin{figure}[tbp]
\includegraphics[bb=69 254 559 558, width=8 cm, clip]{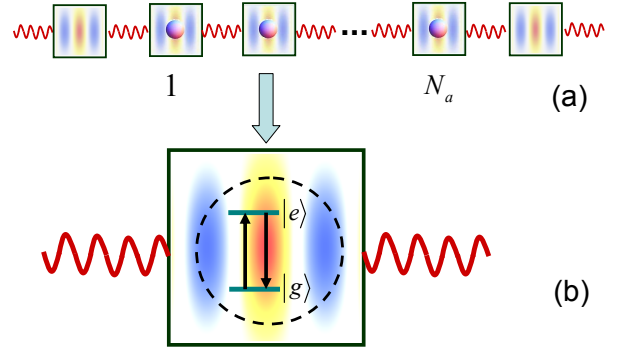}
\caption{(Color online) Schematic setup of the wide-band atomic mirror
model. It is constituted by a coupled-resonator waveguide and an array of
two-level atoms from $1$ to $N_{a}$ as shown in (a). The coupling between
the cavity field and the two-level atom is shown in detail in (b), which is
described by the Jaynes-Cumming model. The existence of the $N_{a}$ atoms
can extend the perfect reflection region to a wide band. }
\end{figure}
Here, $a_{j}$ is the annihilation operator of the $j$th single-mode cavity
with frequency $\omega $, and $V$ is the hopping constant between the
nearest neighbor cavities for the photon. We assume that all the two-level
atoms and the cavity fields have the same energy level spacing $\Omega $ and
frequency $\omega ,$ respectively. The coupling between each atom and the
corresponding cavity field is described by the Jaynes-Cummings model \cite%
{j-c} with homogeneous coupling constant $g$. In Eq. (\ref{3}), the Pauli
spin matrices $\sigma _{j}^{z}\equiv \left\vert e\right\rangle
_{jj}\left\langle e\right\vert -\left\vert g\right\rangle _{jj}\left\langle
g\right\vert $ depicts the atomic energy of the $j$th atom with the ground
state $\left\vert g\right\rangle _{j}$ and exited state $\left\vert
e\right\rangle _{j}$, and $\sigma _{j}^{+}\equiv \left( \sigma
_{j}^{-}\right) ^{\dag }=\left\vert e\right\rangle _{jj}\left\langle
g\right\vert $.

We note that the dispersion relation in the tight-binding boson model in Eq.
(\ref{2}) is the cosine-type, which results in bound states in the hybrid
system. Actually, this tight-binding model is quite appropriate for
simplifying the physical problem and has inspired extensive interests and
lots of attentions both theoretically and experimentally \cite%
{zhou,zhou3,green,fan1,fan2,zhou6,hu}. In our previous work \cite{zhou}, we
have discovered that the electromagnetic field confined in this
coupled-resonator waveguide can be well-controlled by a single two-level
system.

To analyze the transport features of a single photon, we apply the discrete
coordinate scattering approach \cite{zhou,zhou5} by assuming the eigenstate
of $H$ with eigen-energy $E$ for the incident photon in single excitation
subspace as%
\begin{equation}
\left\vert \Psi \left( E\right) \right\rangle
=\sum_{j=-N}^{N}u_{j}^{g}\left\vert j\right\rangle \otimes \left\vert
G\right\rangle +\left\vert 0\right\rangle \otimes
\sum_{j=1}^{N_{a}}u_{j}^{e}\left\vert e\right\rangle _{j}\otimes \left\vert
G_{j}^{\prime }\right\rangle ,  \label{4}
\end{equation}%
where $\left\vert 0\right\rangle $ represents the vacuum of the cavity
fields, $\left\vert j\right\rangle =a_{j}^{\dag }\left\vert 0\right\rangle $%
, and

\begin{equation}
\left\vert G\right\rangle =\prod_{j=1}^{N_{a}}\left\vert g\right\rangle
_{j},\left\vert G_{j}^{\prime }\right\rangle =\prod_{l=1,l\neq
j}^{N_{a}}\left\vert g\right\rangle _{l},
\end{equation}%
Here, $u_{j}^{g}$ and $u_{j}^{e}$\ are the amplitudes of the single photon
and the atomic population in the $j$th cavity, respectively. The first term
on the right hand side of Eq. (\ref{4}) depicts the single-photon
propagating along the waveguide, while the second one represents that the
photon is \textquotedblleft captured" by an atom. It follows from the Schr%
\"{o}dinger equation $H\left\vert \Psi \left( E\right) \right\rangle
=E\left\vert \Psi \left( E\right) \right\rangle $ that the scattering
equations for single photon with discrete coordinate representation read%
\begin{equation}
\omega u_{j}^{g}+V\left( u_{j+1}^{g}+u_{j-1}^{g}\right) +W\left( E\right)
u_{j}^{g}=Eu_{j}^{g}.  \label{5}
\end{equation}%
Here, \ the effective potential%
\begin{equation}
W\left( E\right) =w\left( E\right) \sum_{l=1}^{N_{a}}\delta _{jl}
\end{equation}%
is like a local resonant Dirac-comb \cite{flugge} with strength $w\left(
E\right) =g^{2}/(E-\Omega )$. The equations related to the atomic population
are%
\begin{equation}
\Omega u_{j}^{e}+gu_{j}^{g}=Eu_{j}^{e}.
\end{equation}

We indicate here that $u_{j}^{g}$ and $u_{j}^{e}$\ in Eq. (\ref{4}) depend
on the energy $E$ of the incident photon, and the interaction between the
cavity fields and the atoms provides the potential $W\left( E\right) $ to
affect the propagation of the single photon.\ Eq. (\ref{5}) shows that due
to the array of atoms, the incident single photon acquires an additional
potential described by the local resonant Dirac comb, and the strength of
this potential, i.e., $w\left( E\right) $, depends on the incident photon's
energy $E$. On resonance, i.e., $E=\Omega $, the strength $w\left( E\right) $
of the effective potential is infinite, which definitely leads to the
perfect reflection of the incident photon. This result is consistent with
that in the previous work \cite{zhou}. For the scattering in one dimension,
in which the eigenfunction only possesses the reflection and transmission
waves, the solutions to Eq. (\ref{5}) for $j\neq $ $1$, $2$, ..., $N_{a}$
are supposed to be%
\begin{equation}
u_{j}^{g}=\left\{
\begin{array}{c}
e^{ikj}+re^{-ikj}\text{, \ \ }j<1 \\
te^{ikj}\text{, \ \ }j>N_{a}%
\end{array}%
\right. ,  \label{6a}
\end{equation}%
where $r$ and $t$ are reflection and transmission amplitudes respectively.
To consider elastic scattering we can use the eigenvalue of the scattered
photon
\begin{equation}
E\left( k\right) =\omega +2V\cos k
\end{equation}%
with the cosine-type dispersion for the incident photon with momentum $k$.

The solutions in the region where the cavity fields interact with the atoms
are supposed to be%
\begin{equation}
u_{j}^{g}=r^{\prime }e^{-ik^{\prime }j}+t^{\prime }e^{ik^{\prime }j},\text{ }
\label{6}
\end{equation}%
for $j=1$, $2,...,$ $N_{a}$, where $k^{\prime }$ is the solution of the
transcendental equation%
\begin{equation}
2V\cos k^{\prime }=2V\cos k-w\left( E\right) ,  \label{10}
\end{equation}%
which exhibits the conservation of energy. In Eqs. (\ref{6a}) and (\ref{6}),
$r$ and $t$, together with $r^{\prime }$ and $t^{\prime }$ are determined
below by four boundary conditions, i.e., the scattering equations (\ref{5})
in four points $j=0$, $1$, $N_{a}$, $N_{a}+1$. Then the transmission
coefficient $\left\vert t\right\vert ^{2}$\ is obtained as
\begin{equation}
\left\vert t\right\vert ^{2}=\left\vert \frac{4V^{2}\sin k\sin k^{\prime }}{%
A(E)^{2}e^{ik^{\prime }\left( N_{a}-1\right) }-B(E)^{2}e^{-ik^{\prime
}\left( N_{a}-1\right) }}\right\vert ^{2},  \label{9}
\end{equation}%
corresponding to the reflection coefficients $\left\vert r\right\vert
^{2}=1-\left\vert t\right\vert ^{2},$ where%
\begin{equation}
A(E)=Ve^{-ik^{\prime }}-Ve^{-ik}+w\left( E\right) ,
\end{equation}%
and%
\begin{equation}
B(E)=Ve^{ik^{\prime }}-Ve^{-ik}+w\left( E\right) ,
\end{equation}%
are independent of $N_{a}$.

Furthermore, it follows from Eq. (\ref{10}) that, when%
\begin{equation}
\left\vert \cos k-\frac{w\left( E\right) }{2V}\right\vert \geq 1,  \label{11}
\end{equation}%
$k^{\prime }$ is complex or $k^{\prime }=n\pi $ ($n$ is an integer), which
exhibits the photon's probability decaying in the interaction region. When $%
k^{\prime }=n\pi $, Eq. (\ref{9}) gives that $\left\vert t\right\vert ^{2}=0$%
. And when $k^{\prime }$ is complex, it is shown in Eq. (\ref{9}) that $%
\left\vert t\right\vert ^{2}\rightarrow 0$ when $N_{a}\rightarrow +\infty $.
In this case, the array of atoms behaves as a mirror that reflects the light
perfectly.

The transmission spectra $\left\vert t\right\vert ^{2}$ in Eq. (\ref{9})
versus the detuning $\Delta =E\left( k\right) -\Omega $ for different atom
numbers $N_{a}$\ are shown in Figs. 3(a)-3(d). The parameters in these
figures are chosen as $\omega =5g$, $\Omega =6g$, and $V=-g$. These figures
show that as $N_{a}$ increases, the width of the perfect reflection band
near the resonance $\Delta =0$ increases correspondingly and at last reaches
its maximum value with large $N_{a}$. In contrast to the single atom mirror
case with only one specific reflection frequency, such multi-atom mirror can
be used to manipulate the propagation of a practical wave packet, whose
distribution in momentum space is restricted in the wide-band reflection
region.

\begin{figure}[ptb]
\includegraphics[bb=32 287 513 650, width=4 cm, clip]{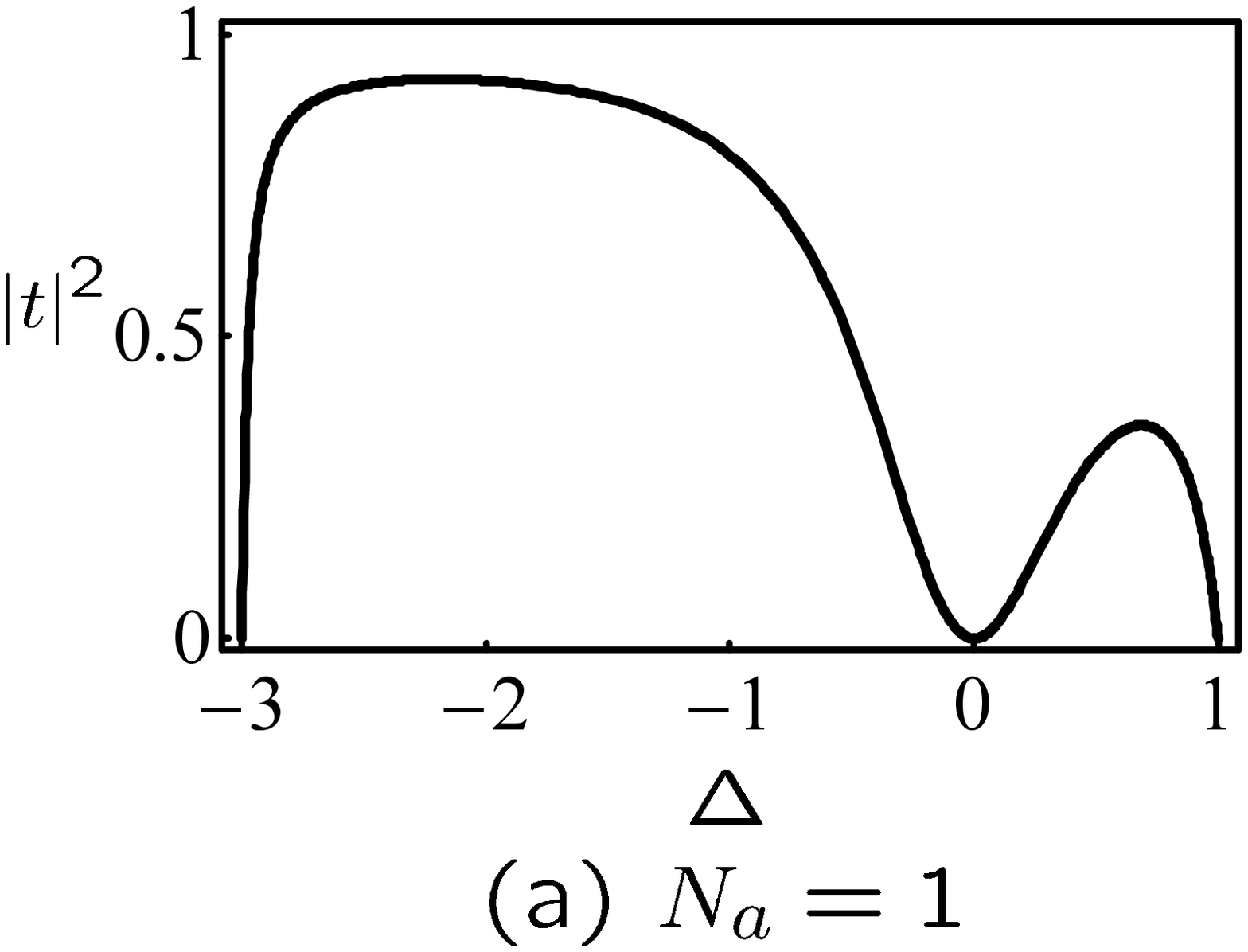} %
\includegraphics[bb=32 287 513 650, width=4 cm, clip]{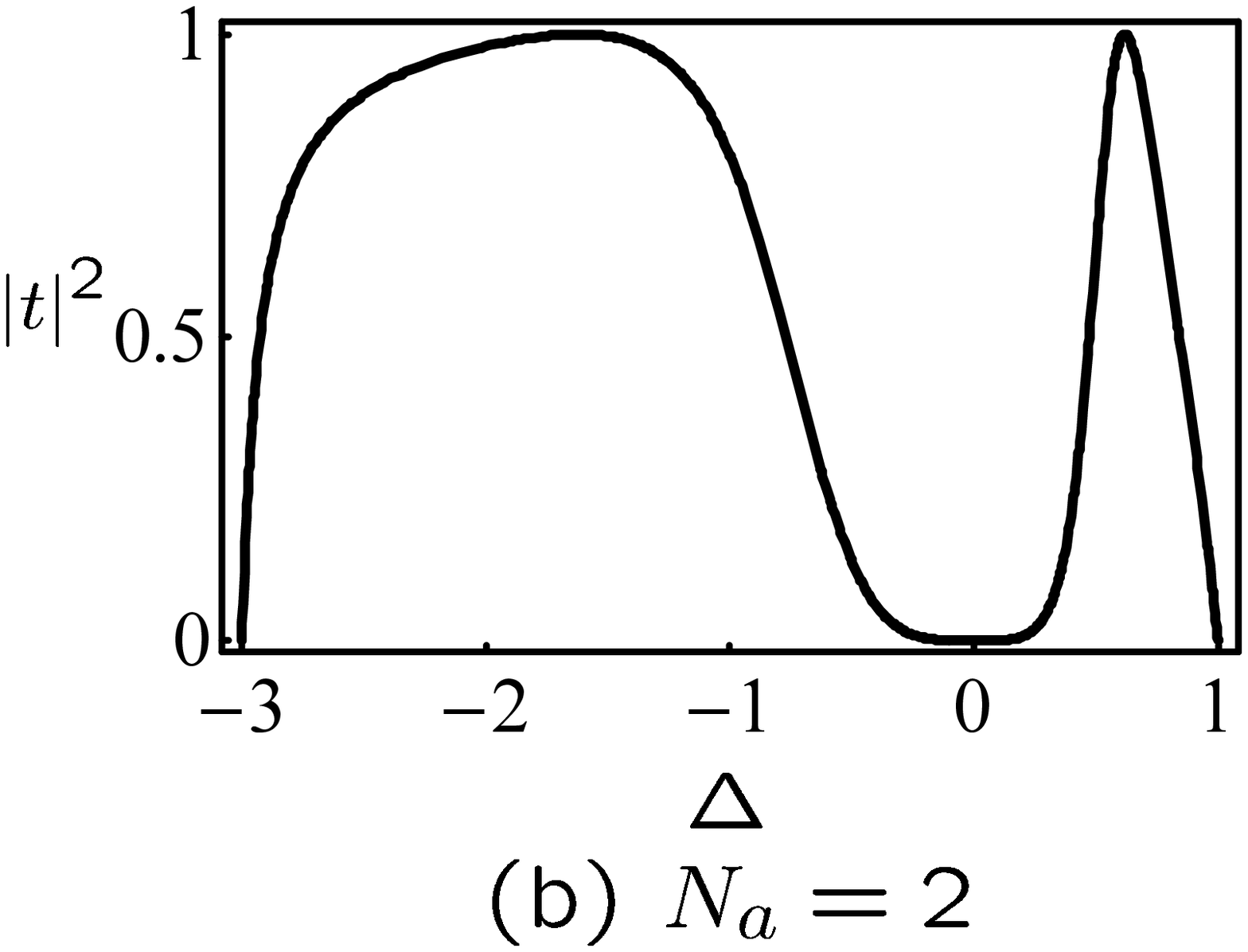} %
\includegraphics[bb=32 287 513 650, width=4 cm, clip]{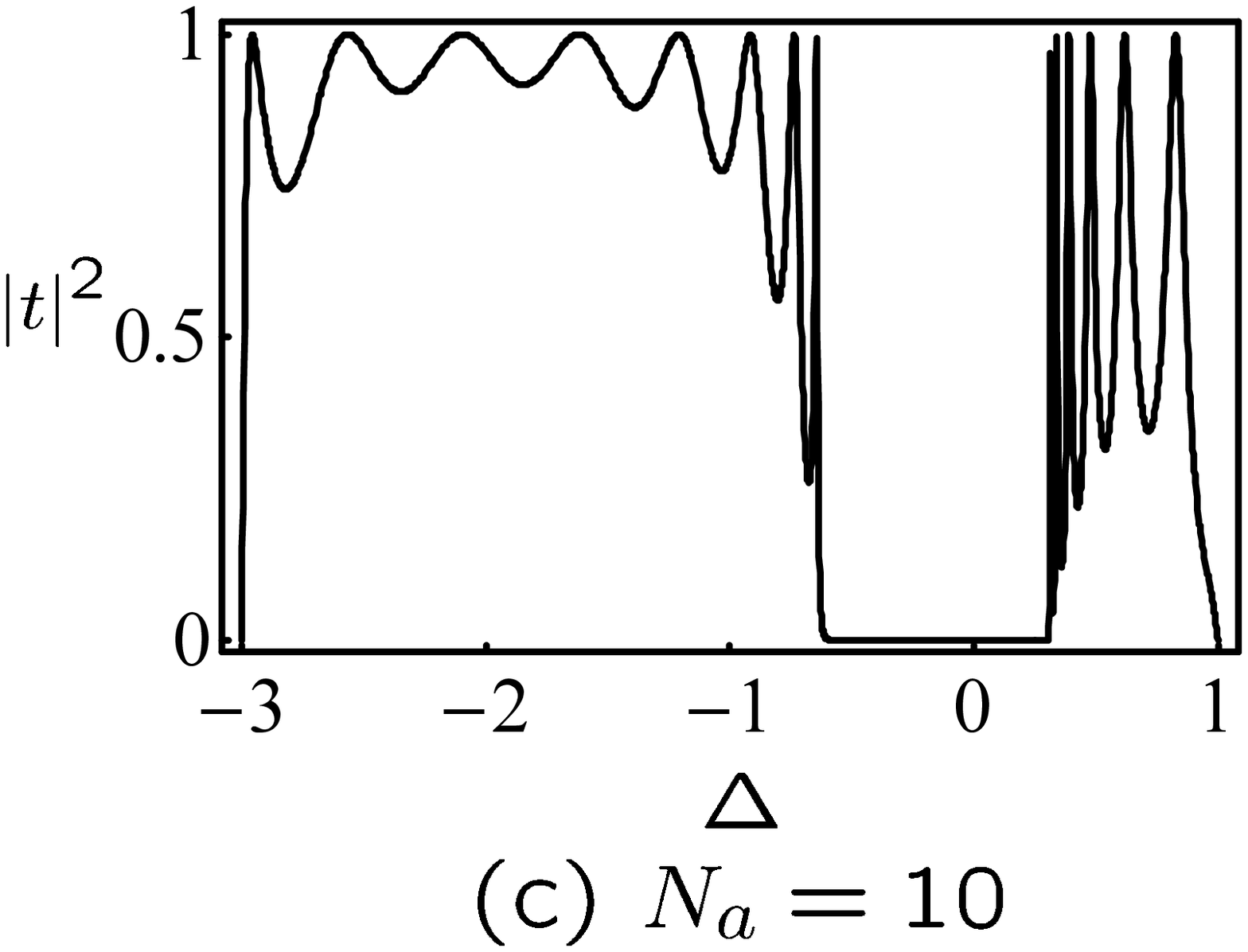} %
\includegraphics[bb=32 287 513 650, width=4 cm, clip]{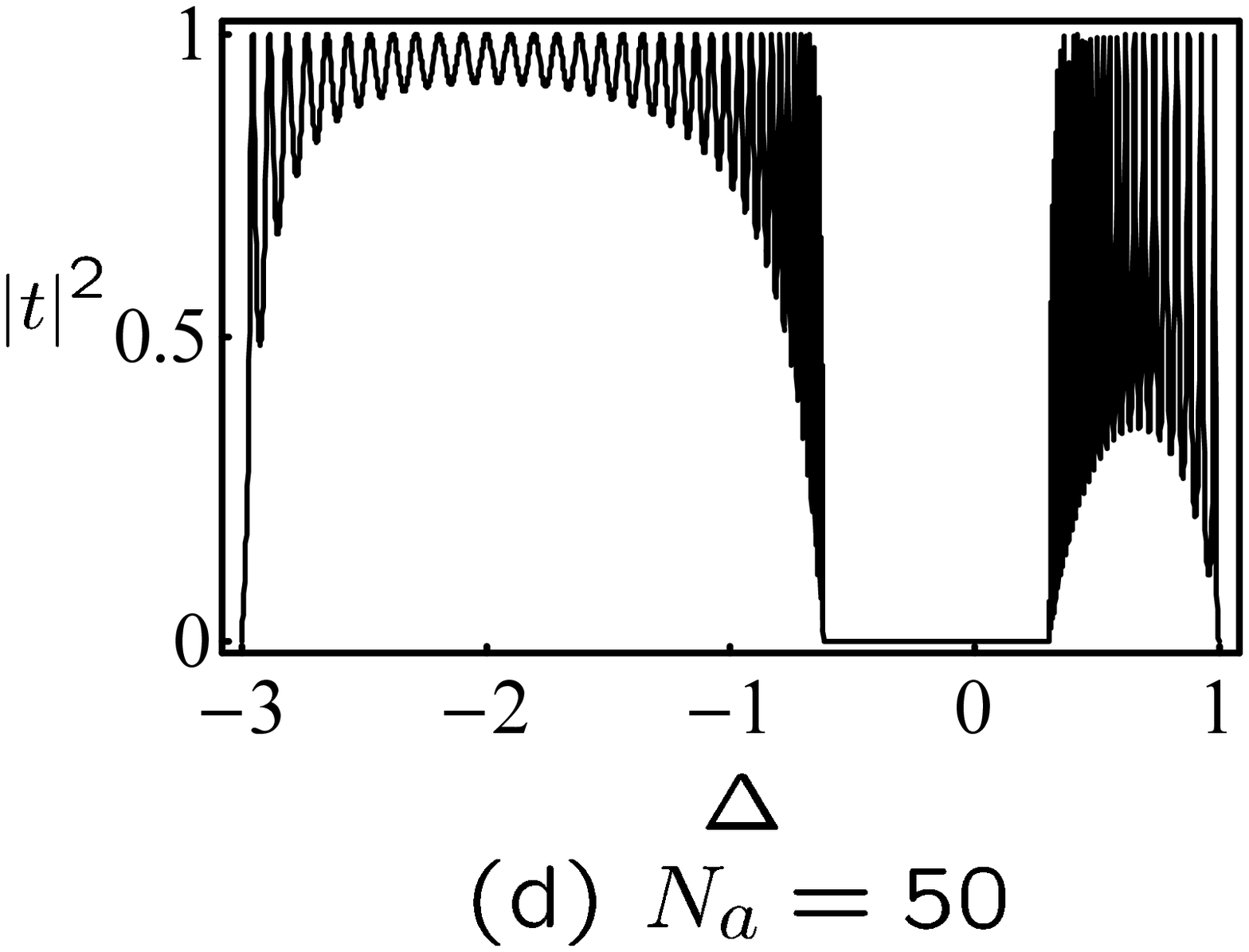}
\caption{The transmission spectrum $\left\vert t\right\vert ^{2}$ as a
function of the detuning $\Delta$ for different atom numbers $N_{a}$, where $%
N_{a}=1$ [for (a)], $N_{a}=2$ [for (b)], $N_{a}=10 $ [for (c)], and $%
N_{a}=50 $ [for (d)]. The wide-band of perfect reflection ($\left\vert
t\right\vert ^{2}=0$) emerges gradually as the atom number $N_{a}$
increases. When doping only a few atoms (like $N_{a}=10$), the near-perfect
reflection band can fit the practical application. When the atom number $%
N_{a}$ tends to infinite, the region of the wide-band reaches its maximum.}
\end{figure}

\section{Frequent Reflections Induced Spectrum Broadening}

We have obtained the wide-band for perfect reflection by solving the
discrete coordinate scattering equations (\ref{5}) in the last section. In
this section, we study the physical mechanism of this wide-band and find the
rigorous boundaries of the band.

As shown in Ref. \cite{zhou}, for single atom at the $0$th single-mode
cavity, the reflection and transmission amplitudes for single incident
photon with momentum $k$ are%
\begin{equation}
r_{1}\left( k\right) =\frac{g^{2}}{-2iV(E-\Omega )\sin k-g^{2}}
\end{equation}%
and%
\begin{equation}
t_{1}\left( k\right) =\frac{2iV(E-\Omega )\sin k}{2iV(E-\Omega )\sin k+g^{2}}%
,
\end{equation}%
respectively. The corresponding normalized eigenstate is%
\begin{equation}
\left\vert \Omega \left( k\right) \right\rangle =\frac{1}{\sqrt{2\pi }}%
\sum_{j=-N}^{N}\xi _{j}^{g}a_{j}^{\dag }\left\vert 0\right\rangle \left\vert
g\right\rangle _{0}+\xi ^{e}\left\vert 0\right\rangle \left\vert
e\right\rangle _{0},  \label{1}
\end{equation}%
where%
\begin{equation}
\xi _{j}^{g}\left( k\right) =\left\{
\begin{array}{c}
e^{ikj}+r_{1}\left( k\right) e^{-ikj},\text{ }j<0 \\
t_{1}\left( k\right) e^{ikj},\text{ }j\geq 0%
\end{array}%
\right. \text{ for }k>0,
\end{equation}%
and%
\begin{equation}
\xi _{j}^{g}\left( k\right) =\left\{
\begin{array}{c}
e^{ikj}+r_{1}\left( -k\right) e^{-ikj},\text{ }j>0 \\
t_{1}\left( -k\right) e^{ikj},\text{ }j\leq 0%
\end{array}%
\right. \text{ for }k<0.
\end{equation}%
These reflection and transmission amplitudes are consistent in magnitudes
with the results we have acquired in Eq. (\ref{9}) when $N_{a}=1$.

An element $S_{kp}$ of the $S$-matrix describing the probability amplitude
of an outgoing photon with momentum $k$ when the incident photon momentum is
$p$ in this single-atom system is%
\begin{equation}
S_{kp}=\delta _{kp}-i2\pi \delta _{E\left( k\right) E\left( p\right)
}\left\langle k\right\vert V_{int}\left\vert \Omega \left( p\right)
\right\rangle ,
\end{equation}%
where%
\begin{equation}
\left\vert k\right\rangle =\frac{1}{\sqrt{2\pi }}%
\sum_{j=-N}^{N}e^{ikj}a_{j}^{\dag }\left\vert 0\right\rangle \left\vert
g\right\rangle _{0}
\end{equation}%
is the outgoing state with momentum $k$,%
\begin{equation}
V_{int}=g\left( a_{0}\sigma _{0}^{+}+a_{0}^{\dag }\sigma _{0}^{-}\right)
\end{equation}%
is the photon-atom coupling. Following these definitions, the $S$-matrix
element reads%
\begin{equation}
S_{kp}=\left\{
\begin{array}{c}
t_{1}\left( p\right) \delta _{kp}+r_{1}\left( p\right) \delta _{-k,p}\text{,
}p>0 \\
t_{1}\left( -p\right) \delta _{kp}+r_{1}\left( -p\right) \delta _{-k,p}\text{%
, }p<0%
\end{array}%
\right. .
\end{equation}

Neglecting the interference between the reflection and transmission waves in
the interaction region, the scattering matrix element $S_{k,k}^{\prime }$
corresponding to the transmission amplitude for $N_{a}$ atoms is written
approximately as%
\begin{equation}
S_{k,k}^{\prime }\approx \left( S_{k,k}\right) ^{N_{a}}=t_{1}^{N_{a}}.
\end{equation}

\begin{figure}[tbp]
\includegraphics[bb=34 378 550 644, width=8 cm, clip]{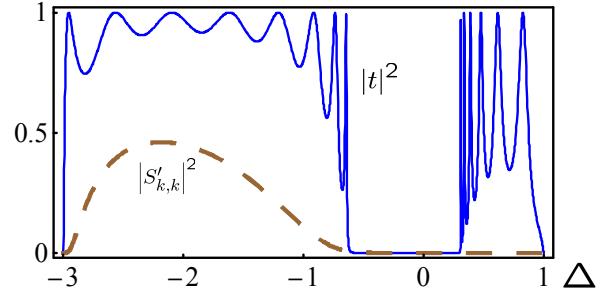}
\caption{(Color online) The transmission coefficient $\left\vert
S_{k,k}^{\prime }\right\vert ^{2}\equiv \left\vert t_{1}\right\vert
^{2N_{a}} $ (brown dashed line) with respect to the detuning $\Delta $ in
the approximation shown in Eq. (25) with $N_{a}=10$. The other parameters
are the same as that in Fig. 3. For comparison, the exact result $\left\vert
t\right\vert ^{2}$ (blue solid line) is plotted in this figure. It is shown
that perfect reflection wide-bands appear in both $\left\vert
S_{k,k}^{\prime }\right\vert ^{2}$ and $\left\vert t\right\vert ^{2}$, which
are due to the reflection of light by multi-atoms. The widths of these two
band are different, as we do not take account of the interference of the
reflection and transmission waves between different atoms in $\left\vert
S_{k,k}^{\prime }\right\vert ^{2}$. Furthermore, the interference effect
leads to the resonant transmission peaks in the transmission spectrum.}
\end{figure}

To investigate this approximation condition, we expand $\left\vert
t\right\vert ^{2}$ shown in Eq. (\ref{9}) around the point $\Delta =0$ as%
\begin{equation}
\left\vert t\right\vert ^{2}\approx \left( \frac{V}{g^{2}}\right) ^{2N_{a}}%
\frac{4V^{2}-\delta ^{2}}{V^{2}}\Delta ^{2N_{a}}+o(\Delta ^{2N_{a}}),
\label{ap1}
\end{equation}%
where we have made an approximation that $\left\vert 2V\cos k\right\vert \ll
\left\vert w\left( E\right) \right\vert $. With this expansion (\ref{ap1}),
we expand $\left\vert t_{1}\right\vert ^{2N_{a}}$ approximately as%
\begin{equation}
\left\vert t_{1}\right\vert ^{2N_{a}}\approx \left( \frac{4V^{2}-\delta ^{2}%
}{g^{4}}\right) ^{N_{a}}\Delta ^{2N_{a}}+o(\Delta ^{2N_{a}}).  \label{ap2}
\end{equation}%
It is shown in Eqs. (\ref{ap1}) and (\ref{ap2}) that in the region near the
resonance, both $\left\vert t\right\vert ^{2}$ and $\left\vert
t_{1}\right\vert ^{2N_{a}}$ tend to zero with the power-law function $\Delta
^{2N_{a}}$. Thus, it indicates that the emergence of the wide band near the
resonance is due to the incoherent reflection by the array of atoms.

This approximate transmission coefficient is plotted in Fig. 4 in contrast
with the rigorous solution shown in Fig. 3(c). Apparently, the wide-band
perfect reflection phenomenon also appears in the approximate transmission
coefficient. It demonstrates that the appearance of the wide-band is due to
the reflection of light by the array of atoms. In other words, when the
light reaches the first cavity coupled to a doped atom, the incident photon
only has the probability $\left\vert t_{1}\right\vert ^{2}$ to pass through
the atom. And then, the second atom repeats this reflection process when the
light passes through it and the transmission coefficient becomes $\left\vert
t_{1}\right\vert ^{4}$. This process is repeated by $N_{a}$ atoms when the
light passes through, and eventually leads to the transmission coefficient
as $\left\vert t_{1}\right\vert ^{2N_{a}}$. In this discussion, we do not
take account of the interference effect at all, as the construction of the
wide-band is only dominated by individual reflection processes. Actually,
when considering the interference of reflection and transmission waves
between atoms, the region forbidding light propagation varies greatly. The
interference effect also leads to the resonate transmission peaks in the
transmission spectrum (see the peaks in Fig. 4).

The width of the perfect reflection band is determined by taking the
imaginary part of the momentum $k^{\prime }$ nonzero or $k^{\prime }=n\pi $.
We notice that in Eq. (\ref{10}), when
\begin{equation}
\cos k^{\prime }=\frac{E-\omega -w\left( E\right) }{2V}\geq 1,  \label{c1}
\end{equation}%
the wave vector $k^{\prime }$ is complex:%
\begin{equation}
k_{+}^{\prime }=2n_{+}\pi +i\alpha _{+}.  \label{12}
\end{equation}%
On the other hand, when%
\begin{equation}
\cos k^{\prime }=\frac{E-\omega -w\left( E\right) }{2V}\leq -1,  \label{c2}
\end{equation}%
$k^{\prime }$ takes the form%
\begin{equation}
k_{-}^{\prime }=\left( 2n_{-}+1\right) \pi +i\alpha _{-}.  \label{13}
\end{equation}%
Here, $n_{\pm }$ are integers and $\alpha _{\pm }$ are real. If $k^{\prime }$
has the form shown in Eq. (\ref{12}) or (\ref{13}), the denominator in Eq. (%
\ref{9}) tends to infinity while $N_{a}\ $is sufficiently large, which
results in the vanishing transmission coefficient.

It follows from Eqs. (\ref{c1}) and (\ref{c2}) that the range for the
incident photon energy $E$ is%
\begin{equation}
\max \{E_{-},E_{\min }\}\leq E\leq \min \{\Omega ,E_{\max }\},  \label{14}
\end{equation}%
or%
\begin{equation}
\max \{\Omega ,E_{\min }\}\leq E\leq \min \{E_{+},E_{\max }\},  \label{15}
\end{equation}%
where%
\begin{equation}
E_{\pm }=\frac{1}{2}(\omega +\Omega )\mp \left\vert V\right\vert \pm \sqrt{%
\left( \frac{\delta }{2}\mp \left\vert V\right\vert \right) ^{2}+g^{2}},
\end{equation}%
\begin{equation}
E_{\max }=\omega +2\left\vert V\right\vert ,E_{\min }=\omega -2\left\vert
V\right\vert ,
\end{equation}%
and%
\begin{equation}
\delta =\omega -\Omega .
\end{equation}%
When $E$ is in this range determined by Eqs. (\ref{14}) and (\ref{15}), with
lots of atoms, the photon is reflected perfectly.

\begin{figure}[ptb]
\includegraphics[bb=44 368 537 597, width=8 cm, clip]{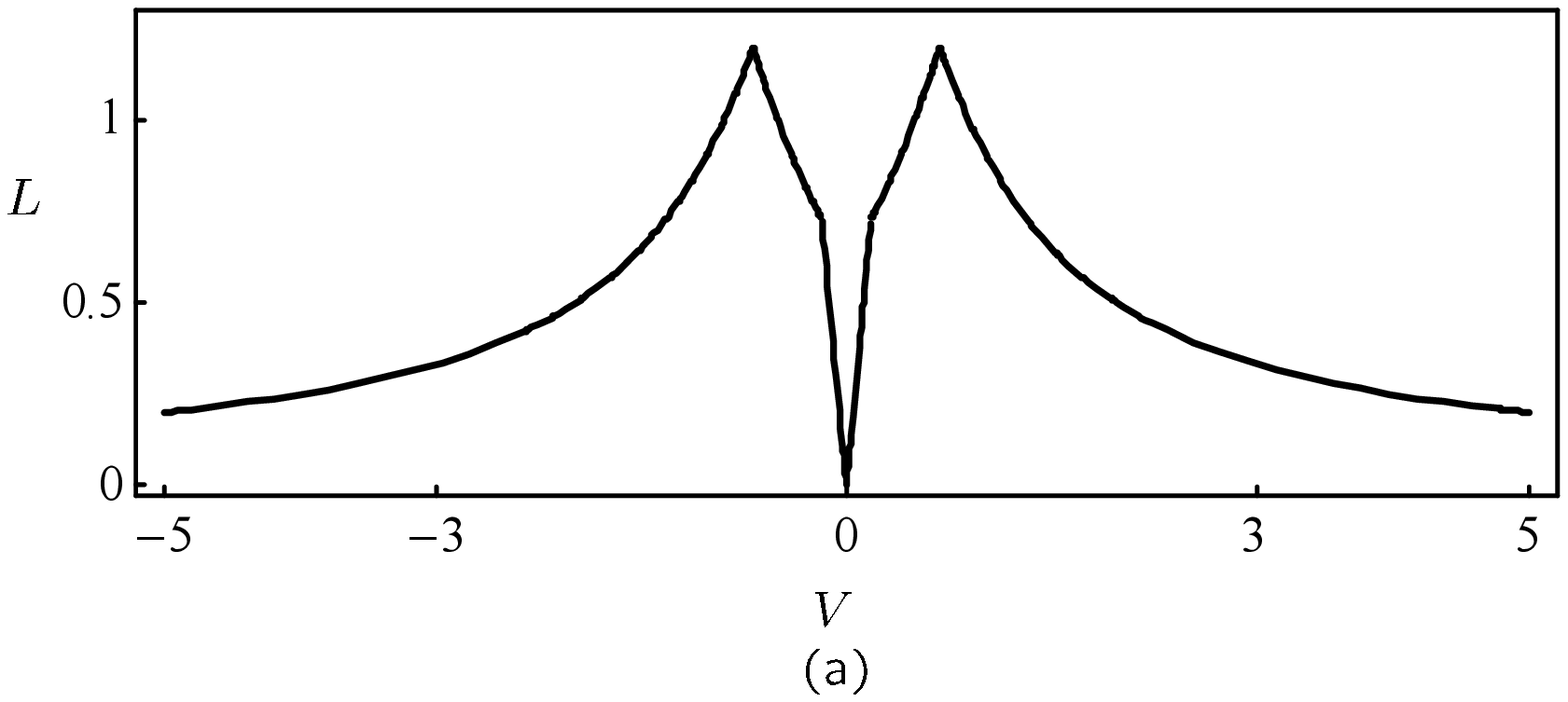} %
\includegraphics[bb=38 355 523 591, width=8 cm, clip]{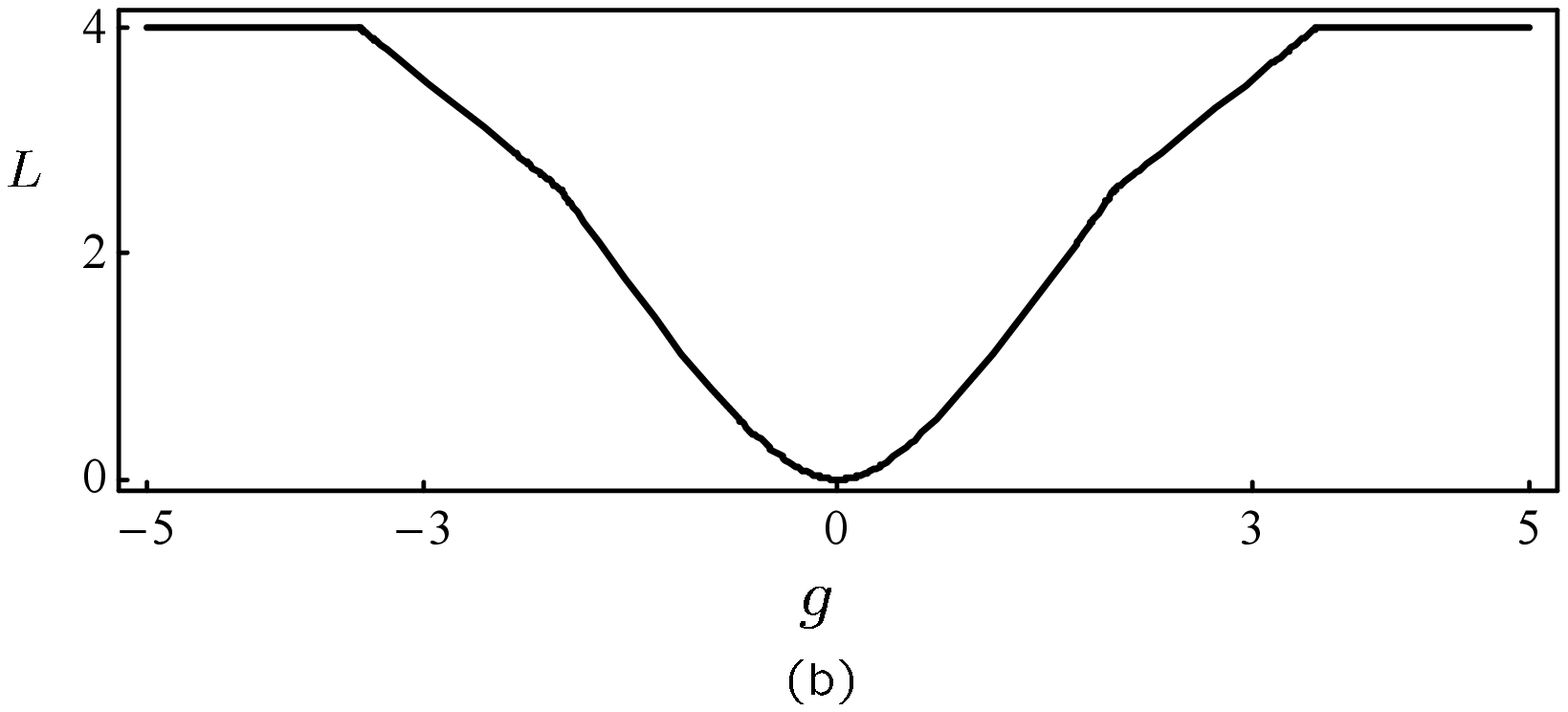} %
\includegraphics[bb=50 314 533 678, width=4 cm, clip]{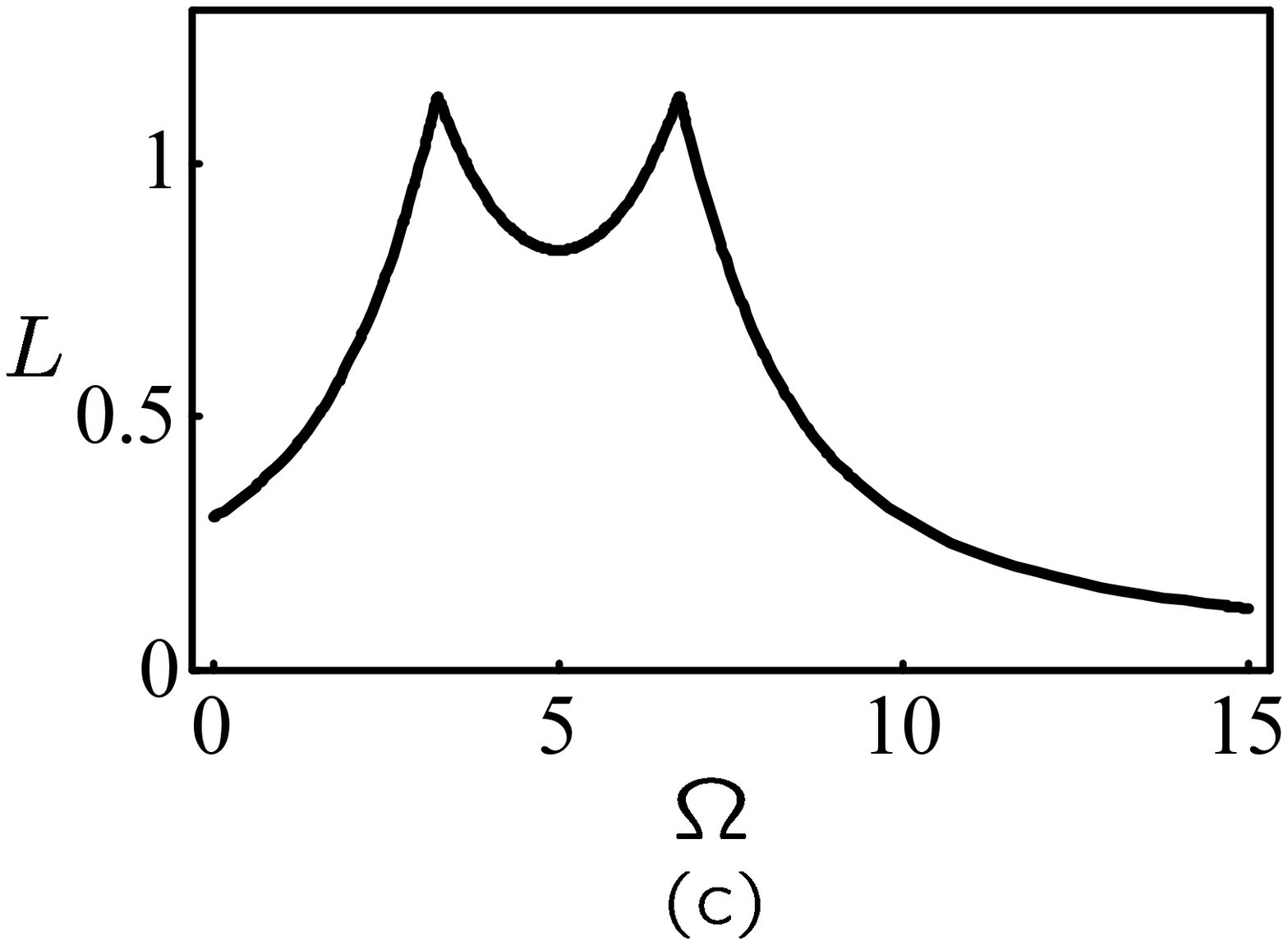} %
\includegraphics[bb=50 314 533 678, width=4 cm, clip]{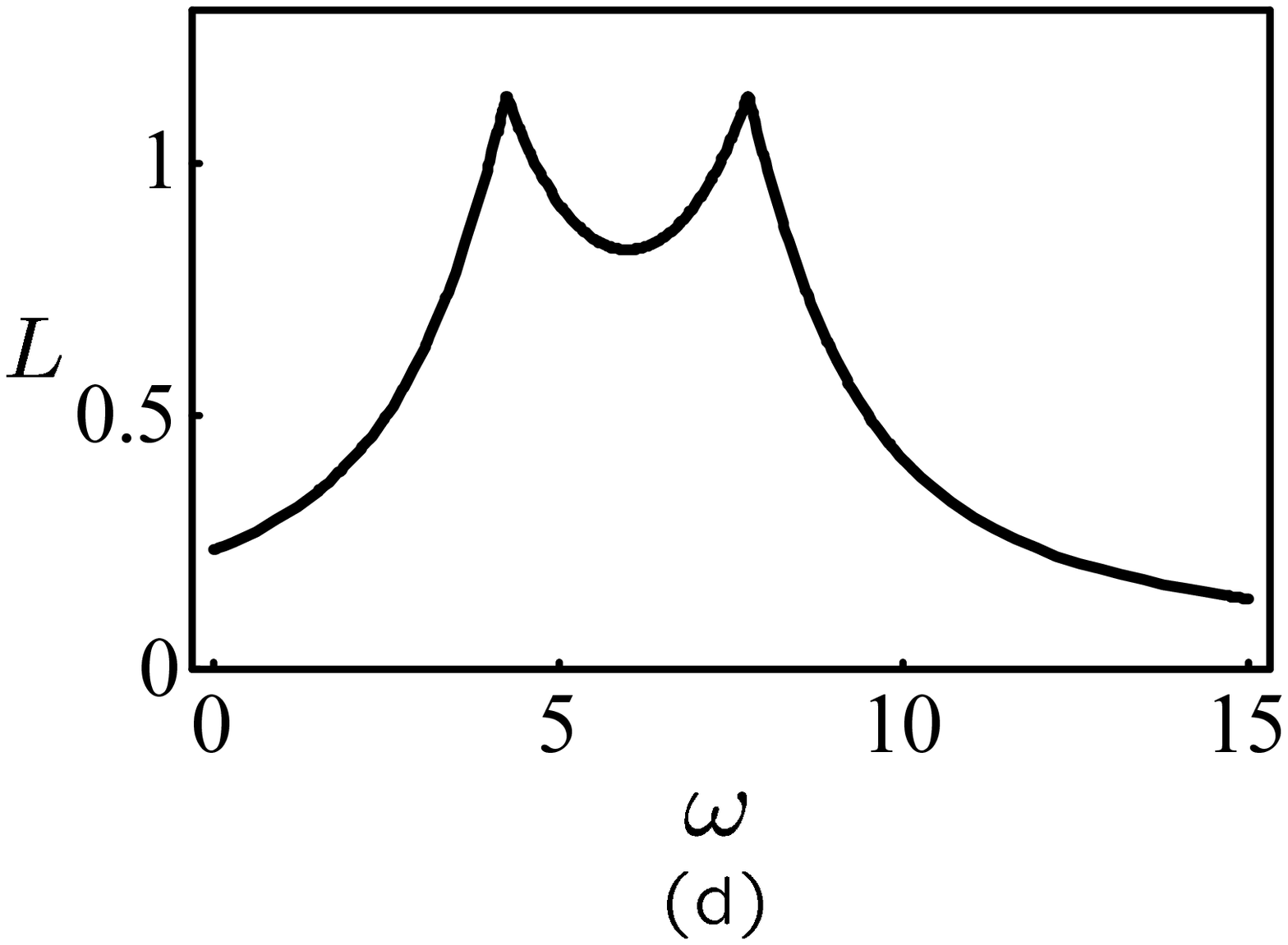}
\caption{The width $L$ for the perfect reflection band as a function of the
parameters $V$ [for (a)], $g$ [for (b)], $\Omega$ [for (c)], and $\protect%
\omega$ [for (d)], with other parameters the same as that in Fig. 3. In Fig.
5(a), $L$ firstly increases and then decreases with the increment of $%
\left\vert V\right\vert$. In Fig. 5(b), the width $L$ increases as the
atom-photon coupling strength $\left\vert g\right\vert $ increases, and $L$
reaches its maximum $4 \left\vert V\right\vert$ at last. In Figs. 5(c) and
5(d), in the region where the difference between the incident photon's
energy and the atom's energy level spacing is very large, the width $L$
tends to zero.}
\end{figure}

When $g=0$, the wide-band width $L=0$, which corresponds to our common sense
that the photon propagates free along the CRW without coupling to atoms.
With the set of parameters in Fig. 3, the width $L$ is $E_{+}-E_{-}$, with
the resonate point $\Delta =0$ in the wide band.\ But note that with some
other parameters, the point $\Delta =0$ does not locate inside the
wide-band. Additionally, when parameters satisfy one of the conditions as
follows: (i) $E_{-}\leq E_{\min }\leq \Omega \leq E_{\max }\leq E_{+}$; (ii)
$E_{\min }\geq \Omega $ and $E_{+}\geq E_{\max }$; (iii) $E_{\min }\leq
\Omega $ and $E_{-}<E_{\min }$, the band width is $4\left\vert V\right\vert $%
. Namely, by tuning the parameters in this region, the light can be
reflected perfectly in the whole region of the energy of the incident
photon. With other parameters the same as that in Fig. 3, we plot the
wide-band width $L$ with respect to the parameters $V$, $g$, $\Omega $, $%
\omega $ respectively in Figs. 5(a)-5(d). It is shown in Fig. 5(a) that when
$V=0$, which means that the photon can not hop in the CRW, $L=0$. As $%
\left\vert V\right\vert $ increases, $L$ increases until $\left\vert
V\right\vert $ reaches some critical point. Then $L$ decreases when $%
\left\vert V\right\vert $ increases, since in this range, the larger the
hopping strength $\left\vert V\right\vert $ is, the weaker the photon-atom
coupling becomes as a perturbation. In Fig. 5(b), $L$ is a monotonic
increasing function of the atom-photon coupling strength $\left\vert
g\right\vert $, and at last $L$ reaches the maximum $4\left\vert
V\right\vert $. Namely, the stronger coupling leads to a wider perfect
reflection band. Eqs. 5(c) and 5(d) exhibit that when $\Omega $ or $\omega $
is very large, i.e., the difference between the incident photon's energy $%
E\left( k\right) $ and the two-level atom's energy level spacing $\Omega $
is very large, $L$ tends to zero. The reason leading to this behavior is
similar to that in the rotating-wave approximation \cite{j-c}, i.e., the
large energy difference makes the interaction negligible with large time
scale.

The width of the perfect reflection band is obtained with large $N_{a}$. We
now discuss how large $N_{a}$ is to insure our result reliable. With the
parameters in Fig. 3, the boundaries $\Delta _{\pm }$ of the wide-band are
acquired using Eqs. (\ref{14}) and (\ref{15}) as%
\begin{equation}
\Delta _{-}\approx -0.618g\text{ and }\Delta _{+}\approx 0.302g.
\end{equation}

\begin{figure}[tbp]
\includegraphics[bb=29 315 547 607, width=8 cm, clip]{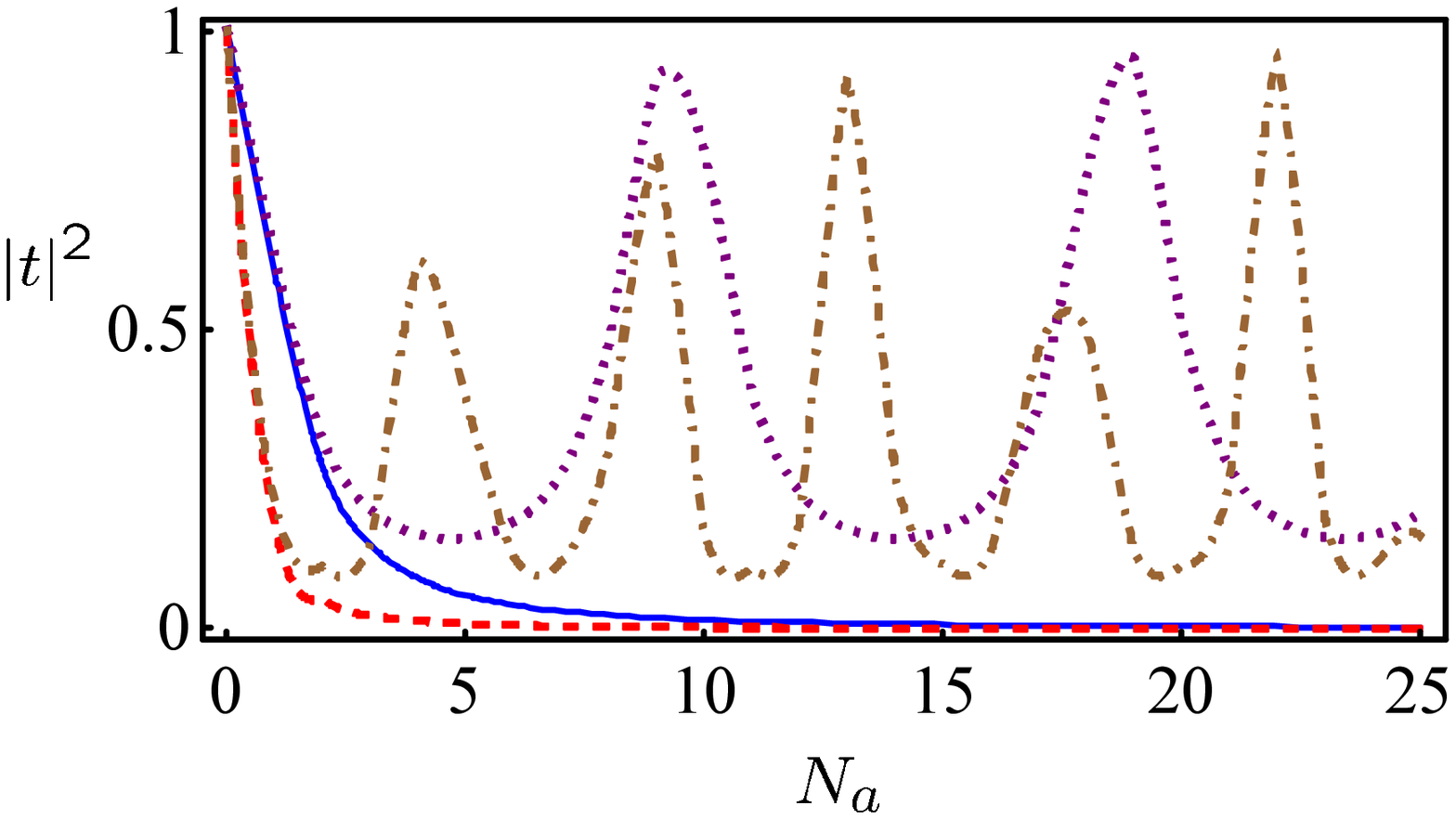}
\caption{(Color online) The transmission coefficients $\left\vert
t\right\vert ^{2}$ in two boundaries $\Delta _{\pm }$ with respect to $N_{a}$%
, where the blue dotted line represents $\left\vert t\right\vert ^{2}$ in
the left boundary, while the red dashed line represents $\left\vert
t\right\vert ^{2}$ in the right boundary. When $N_{a}\geq 20$, the
transmission coefficients in both the two boundaries vanish approximately.
So with this set of parameters, $20$ is large enough for $N_{a}$ to insure
the reliability for the width of the perfect reflection band we obtained.
For comparison, the transmission coefficients with the energy beyond the
wide band are plotted (the purple dotted line for $\Delta =-0.65$, and the
brown dotdashed line for $\Delta =0.35$). The transmission coefficients $%
\left\vert t\right\vert ^{2}$ does not tends to zero as the atom number $%
N_{a}$ increases unless the energy is in the wide band.}
\end{figure}
We plot the transmission coefficients $\left\vert t\right\vert ^{2}$ in
these two boundaries with respect to $N_{a}$ in Fig. 6. When $N_{a}\geq 20$,
the transmission coefficients in both boundaries vanish approximately. As a
result, the width we obtained of perfect reflection does not require $%
N_{a}\rightarrow \infty $ in practice. In fact, with this set of parameters
in Fig. 6, $N_{a}=20$ is large enough to build the wide-band for perfect
reflection. For comparison, we also plot $\left\vert t\right\vert ^{2}$
versus $N_{a}$ with the energies beyond the wide band, which show that $%
\left\vert t\right\vert ^{2}$ does not tend to zero with the increment of $%
N_{a}$, even the energy is very close to that at the boundaries $\Delta
_{\pm }.$

\section{Slow Light Resonant Absorption}

The wide-band reflection by multi-atom mirror has been shown above. Now we
consider the slowing and stopping light phenomena in the interaction region
with an array of atoms, which actually has a close relation to the emergence
of the wide band \cite{hu}.

We revisit the Hamiltonian in the interaction region
\begin{align}
H_{int}& =\sum_{j=1}^{N_{a}}\omega a_{j}^{\dag
}a_{j}+V\sum_{j=1}^{N_{a}-1}\left( a_{j}^{\dag }a_{j+1}+a_{j+1}^{\dag
}a_{j}\right)  \notag \\
& +\sum_{j=1}^{N_{a}}\left[ \frac{\Omega }{2}\left( \sigma _{j}^{z}+1\right)
+g\left( a_{j}\sigma _{j}^{+}+a_{j}^{\dag }\sigma _{j}^{-}\right) \right] .
\label{a}
\end{align}%
The corresponding eigenstates for Eq. (\ref{a}) is%
\begin{equation}
\left\vert \Psi _{int}\right\rangle =\sum_{j=1}^{N_{a}}v_{j}^{g}\left\vert
j\right\rangle \otimes \left\vert G\right\rangle +\left\vert 0\right\rangle
\sum_{j=1}^{N_{a}}v_{j}^{e}\left\vert e\right\rangle _{j}\otimes \left\vert
G_{j}^{\prime }\right\rangle ,
\end{equation}%
with eigenvalue $E$ that is determined by the set of equations%
\begin{align}
\omega v_{j}^{g}+V\left( v_{j+1}^{g}+v_{j-1}^{g}\right) +w\left( E\right)
v_{j}^{g}& =Ev_{j}^{g},  \notag \\
j& =2,...,N_{a}-1,  \label{b}
\end{align}%
\begin{equation}
\omega v_{1}^{g}+Vv_{1}^{g}+w\left( E\right) v_{1}^{g}=Ev_{1}^{g},
\label{bc}
\end{equation}%
\begin{equation}
\omega v_{N_{a}}^{g}+Vv_{N_{a}-1}^{g}+w\left( E\right)
v_{N_{a}}^{g}=Ev_{N_{a}}^{g},  \label{c}
\end{equation}%
and%
\begin{equation}
\Omega v_{j}^{e}+gv_{j}^{g}=Ev_{j}^{e}.
\end{equation}

We note the the difference between Eqs. (\ref{b})-(\ref{c}) and Eq. (\ref{5}%
) is the different boundary conditions. The solutions to the set of
equations (\ref{b})-(\ref{c})\ are%
\begin{equation}
v_{j}^{g}=A\sin pj,
\end{equation}%
with normalization constant $A,$\ and the eigenvalues satisfy%
\begin{equation}
E_{int}=\omega +\frac{g^{2}}{E_{int}-\Omega }+2V\cos p,  \label{e}
\end{equation}%
where $p$ is determined by the boundary conditions (\ref{bc}) and (\ref{c})
as%
\begin{equation}
p=\frac{n\pi }{N_{a}+1},n=1,...N_{a}.
\end{equation}%
Consequently,
\begin{equation}
E_{int}^{\pm }\left( p\right) =\frac{1}{2}(\delta _{p}\pm \sqrt{\delta
_{p}^{2}+4g^{2}})+\Omega ,  \label{ee}
\end{equation}%
where $\delta _{p}=\delta +2V\cos p.$

\begin{figure}[ptb]
\includegraphics[bb=147 566 524 764, width=8 cm, clip]{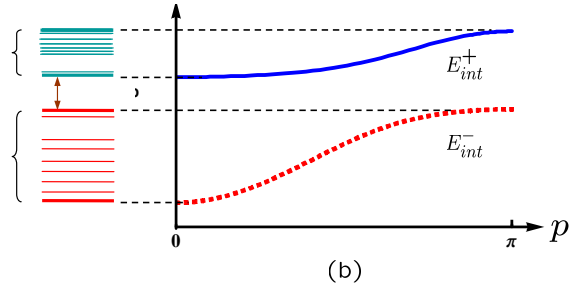}
\caption{(Color online) Schematic configuration for the two energy bands $%
E_{int}^{\pm}$ [for (a)] and figure for $E_{int}^{\pm}\left( p\right) $ with
respect to $p$ [for (b)]. Each band has $N_{a}$ energy levels, and when $%
N_{a} $ energy levels tends to infinity, the band gap is $E^{+}-E^{-}$.}
\end{figure}

It is shown in Eq. (\ref{ee}) that there are two energy bands for the
interaction region, which are labeled by $E_{int}^{+}$ and $E_{int}^{-}$.
The two bands are shown schematically in Fig. 7(a). When $N_{a}$ tends to
infinity, the energy at bottom of the upper band $E_{int}^{+}$ is $E_{+}$,
and at top of the lower band $E_{int}^{-}$ the energy is $E_{-}$, with the
band gap $L=E_{+}-E_{-}$ as shown in Eqs. (\ref{14}) and (\ref{15}). The
band gap for the interaction region contains the wide band for perfect
reflection of single photon. Namely, when the incident photon energy is in
the band gap, it is impossible for the photon to go through the interaction
region. With large $N_{a}$, we plot $E_{int}^{\pm }\left( p\right) $ with
respect to $p$ in Fig. 7(b). Here, the parameters are the same as that in
Fig. 3. We note that the incident electron's energy corresponding to the
resonant peaks in the transmission spectrum are not in agreement with the
eigen-energies we obtained in Eq. (\ref{ee}) in the interaction region.

The group velocity $v_{g}$\ of light propagating in the interaction region
is defined as%
\begin{equation}
v_{g}^{\pm }\left( p\right) =\partial _{p}E_{int}^{\pm }\left( p\right) .
\end{equation}%
Since the momentum $k^{\prime }$ in the interaction region depends on the
energy $E$ of the incident photon, the group velocity $v_{g}$ is also
dependent on $E.$\ However, when $E$ is in the band gap, $k^{\prime }$ is
complex, and $v_{g}\left( \Delta \right) $ is also complex. This complex $%
v_{g}\left( \Delta \right) $ depicts the decay in the light propagating in
the interaction region. In this sense, when $N_{a}$ is large, this group
velocity is zero. It demonstrates that the stopping light phenomenon is due
to the atomic mirror. The group velocities in the two bands are plotted in
Fig. 8 with the same parameters in Fig. 3. For comparison, we also plot the
group velocity $v_{g}=-2V\sin k$ in the CRW without doping atoms. When $%
\Delta <0$, the photon propagates in the lower band, while $\Delta >0$, the
photon propagates in the upper band. Fig. 8 shows the slowing and stopping
light phenomena in the interaction region, and the width of $\Delta $ in the
light stopping region corresponds exactly to the width of the perfect
reflection band shown in Fig. 3(d).

\begin{figure}[ptb]
\includegraphics[bb=71 295 515 619, width=6 cm, clip]{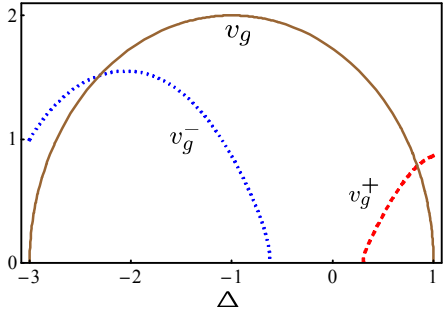}
\caption{(Color online) The light group velocity $v_{g}^{-}$ (dotted blue
line),$v_{g}^{+}$ (dashed red line),and $v_{g}$ (solid brown line) for lower
band $E^{-}$, upper band $E^{+}$, and free tight-binding model $E\left(
k\right) $ versus the detuning $\Delta$. The region where $v_{g}^{\pm}=0$
depicting the stopping light phenomenon exactly corresponds to the wide band
for perfect reflection.}
\end{figure}

We note that these results based on the assumption that $N_{a}$ is large, or
even infinite. However, in practical applications, $N_{a}$ is finite and
maybe not very large, so the reflection in the whole of the band would be
not so perfect. We investigate the influence of $N_{a}$ on the perfect
reflection wide-band, and show that with some parameters, $N_{a}=20$ is
large enough to ensure the the reliability of our results.

\section{Influence of Imperfections in Experimental Implements}

In experiments, the atomic decay, the intrinsic loss of each coupled
resonators, and the disorder \cite{green,dis,dis1} in the CRW, are
inevitable. In this section, we investigate how these imperfections
influence the frequency wide-band for perfect reflection we have studied.
Firstly, we consider the disorder problem without atomic decay and losses of
coupled resonators.

In principle, the whole CRW's scale can be infinite, thus we only consider a
segment of disordered resonators. Moreover, we assume the disordered region
happens to the atom-cavity interaction region, which means the hopping
constant $V_{j}$ for $i$, $j\in \left[ 1\text{, }N_{a}-1\right] $, and the
on-site frequency $\omega _{j}$ for $j\in \left[ 1\text{, }N_{a}\right] $
become position dependent. Under these assumptions, the total Hamiltonian%
\begin{equation}
H_{imp}=H_{L}+H_{int}^{D}  \label{him}
\end{equation}%
is divided into two parts as: the leads' part%
\begin{eqnarray}
H_{L} &=&\omega \left( \sum_{j=-N}^{0}+\sum_{j=N_{a}+1}^{N}\right)
a_{j}^{\dag }a_{j}+ \\
&&V\left( \sum_{j=-N}^{0}+\sum_{j=N_{a}}^{N}\right) \left( a_{j}^{\dag
}a_{j+1}+a_{j+1}^{\dag }a_{j}\right) ,
\end{eqnarray}%
and the disordered part in the atom-cavity interaction region%
\begin{align}
H_{int}^{D}& =\sum_{j=1}^{N_{a}}\omega _{j}a_{j}^{\dag
}a_{j}+\sum_{j=1}^{N_{a}-1}V_{j}\left( a_{j}^{\dag }a_{j+1}+a_{j+1}^{\dag
}a_{j}\right)   \notag \\
& +\sum_{j=1}^{N_{a}}\left[ \frac{\Omega }{2}\left( \sigma _{j}^{z}+1\right)
+g\left( a_{j}\sigma _{j}^{+}+a_{j}^{\dag }\sigma _{j}^{-}\right) \right] .
\end{align}%
The eigenstate of $H_{imp}$ with eigen-energy $E=\omega +2V\cos k$ for the
incident photon with momentum $k$ in single excitation subspace has the
similar form of that in Eq. (\ref{4}) as%
\begin{equation}
\left\vert \Phi \left( E\right) \right\rangle =\left(
\sum_{j=-N}^{0}+\sum_{j=N_{a}+1}^{N}\right) \phi _{j}\left\vert
j\right\rangle \otimes \left\vert G\right\rangle +\left\vert \Phi _{D}\left(
E\right) \right\rangle ,
\end{equation}%
where the amplitudes in leads part's wave-function can be still assumed to
describe the reflection and transmission as
\begin{equation}
\phi _{j}=\left\{
\begin{array}{c}
e^{ikj}+r_{D}e^{-ikj}\text{, \ \ }j<1 \\
t_{D}e^{ikj}\text{, \ \ }j>N_{a}%
\end{array}%
\right. ,
\end{equation}%
and%
\begin{equation}
\left\vert \Phi _{D}\left( E\right) \right\rangle =\sum_{j=1}^{N_{a}}\phi
_{j}^{g}\left\vert j\right\rangle \otimes \left\vert G\right\rangle
+\left\vert 0\right\rangle \otimes \phi _{j}^{e}\left\vert e\right\rangle
_{j}\otimes \left\vert G_{j}^{\prime }\right\rangle
\end{equation}%
is the wave function in the atom-cavity interaction region. Here, $r_{L}$
and $t_{L}$ are the reflection and transmission amplitudes, respectively.
Resulting from the continuous conditions of both the interfaces between the
leads and the atom-cavity interaction region, those amplitudes satisfy%
\begin{equation}
r_{D}=e^{ik}\left( \phi _{1}^{g}-e^{ik}\right) ,
\end{equation}%
\begin{equation}
t_{D}=e^{-ikN_{a}}\phi _{N_{a}}^{g}.
\end{equation}%
By solving the Schrodinger equation $H_{imp}\left\vert \Phi \left( E\right)
\right\rangle =E\left\vert \Phi \left( E\right) \right\rangle $, we
straightforwardly obtain the equation for $\left\vert \Phi _{D}\left(
E\right) \right\rangle $ as
\begin{equation}
\left\vert \Phi _{D}\left( E\right) \right\rangle =V\left( e^{2ik}-1\right)
w\left\vert 1\right\rangle \otimes \left\vert G\right\rangle ,  \label{hh}
\end{equation}%
where%
\begin{equation}
w=\frac{1}{H_{int}^{D}+Ve^{ik}\left( a_{1}^{\dag }a_{1}+a_{N_{a}}^{\dag
}a_{N_{a}}\right) -E}.
\end{equation}%
is the inverse of the Hamiltonian matrix together with the contributions of
leads.

The amplitudes $\phi _{1}^{g}$ and $\phi _{N_{a}}^{g}$ are completely
determined by the $2N_{a}\times 2N_{a}$ matrix $w$. Since the disorder
exists in atom-cavity interaction region, the inhomogeneity of the hopping
constants $\left\{ V_{j}\right\} $ and on-site frequencies $\left\{ \omega
_{j}\right\} $ are expected to destroy the coherence of the incident photon
and hence enhances the reflection. For a particular realization of disorder,
i.e., for given sets $\left\{ \omega _{j}\right\} $ and $\left\{
V_{j}\right\} $ which are generated randomly in the ranges $\left[ \omega
-0.2\omega ,\omega +0.2\omega \right] $ and $\left[ V-0.2V,V+0.2V\right] $
respectively, we plot the transmission coefficient $\left\vert
t_{D}\right\vert ^{2}$ versus the detuning $\Delta =E\left( k\right) -\Omega
$ in Fig. 9(a). Other parameters are the same as that in Fig. 3(c). Fig.
9(a) shows that the wide-band is almost not affected, while out the
wide-band the transmission coefficient is changed dramatically. We assume
the disorder in the atom-cavity interaction region has the Gaussian
distribution of both $\left\{ \omega _{j}\right\} $ and $\left\{
V_{j}\right\} $ as%
\begin{equation}
P(x)=\frac{\exp \left[ (x-x_{0})^{2}/2\sigma ^{2}\right] }{2\pi \sigma },
\end{equation}%
where $P(x)$ is the probability for a given value $x$, $x=\omega _{j}$, $%
V_{j}$, $x_{0}$ is the averaged value, and $\sigma $ is the variance. We
plot the averaged transmission coefficient $\left\vert t_{D}\right\vert ^{2}$
versus $\Delta $ in Fig. 9(b)-(d). Here, for the sets $\left\{ \omega
_{j}\right\} $ and $\left\{ V_{j}\right\} $, $x_{0}=\omega $ and $V$, and
the variances are respectively $0.01\omega $ and $0.01\left\vert
V\right\vert $ for Fig. 9(b), $0.05\omega $ and $0.05\left\vert V\right\vert
$ for Fig. 9(c), $0.1\omega $ and $0.1\left\vert V\right\vert $ for Fig.
9(d). It is shown that Fig. 9(b) is quite similar to Fig. 3(c), both in the
perfect and non-perfect reflection region. When the variances of $\left\{
\omega _{j}\right\} $ and $\left\{ V_{j}\right\} $ increase, the range of
the wide-band is shortened, and in other region, the transmission curve is
smoothed, with the depressed envelop, which demonstrates the destruction of
the coherence of the incident photon due to the existence of the disorder.

\begin{figure}[ptb]
\includegraphics[bb=62 410 520 784, width=4 cm, clip]{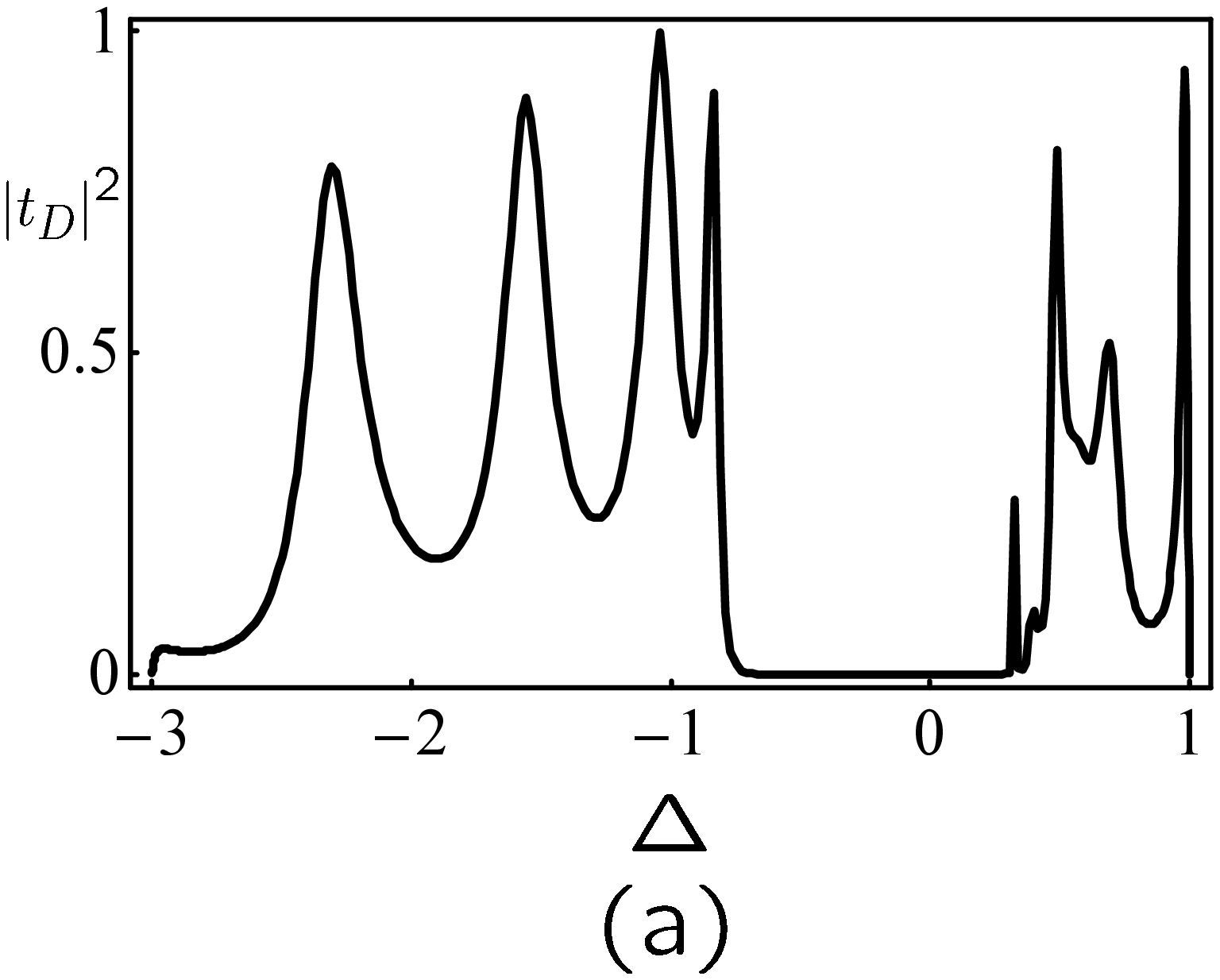} %
\includegraphics[bb=62 410 520 784, width=4 cm, clip]{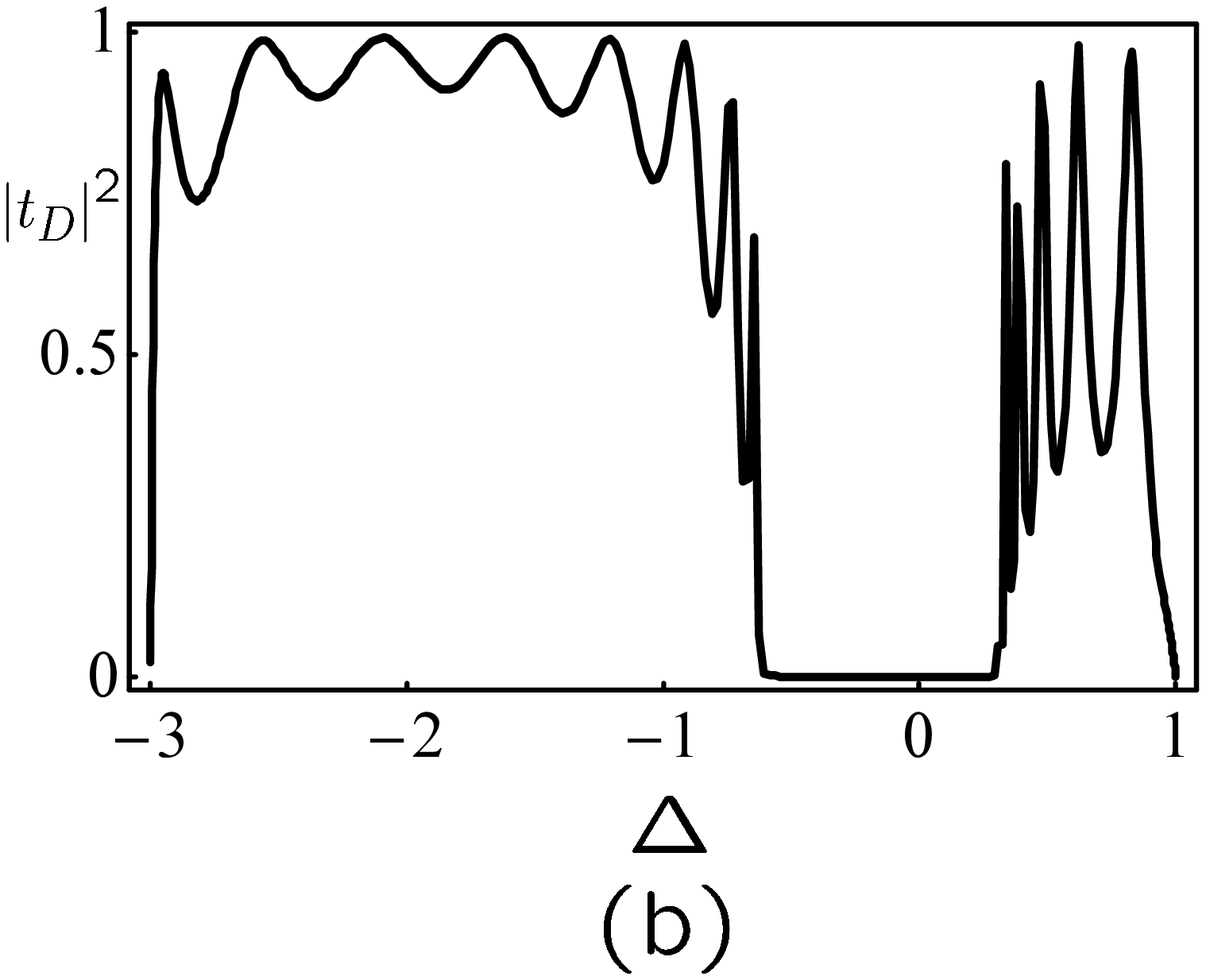} %
\includegraphics[bb=62 410 520 784, width=4 cm, clip]{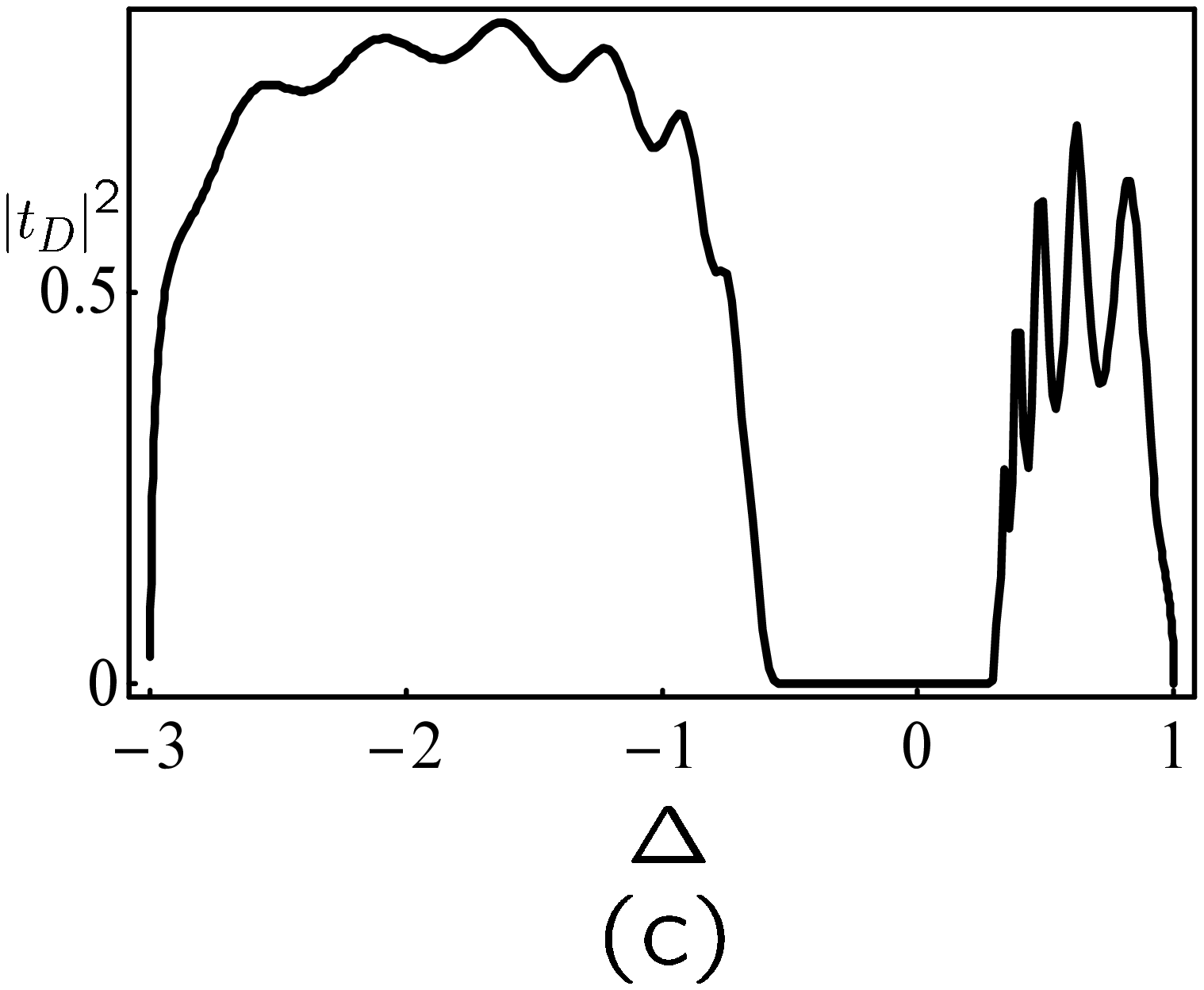} %
\includegraphics[bb=62 410 520 784, width=4 cm, clip]{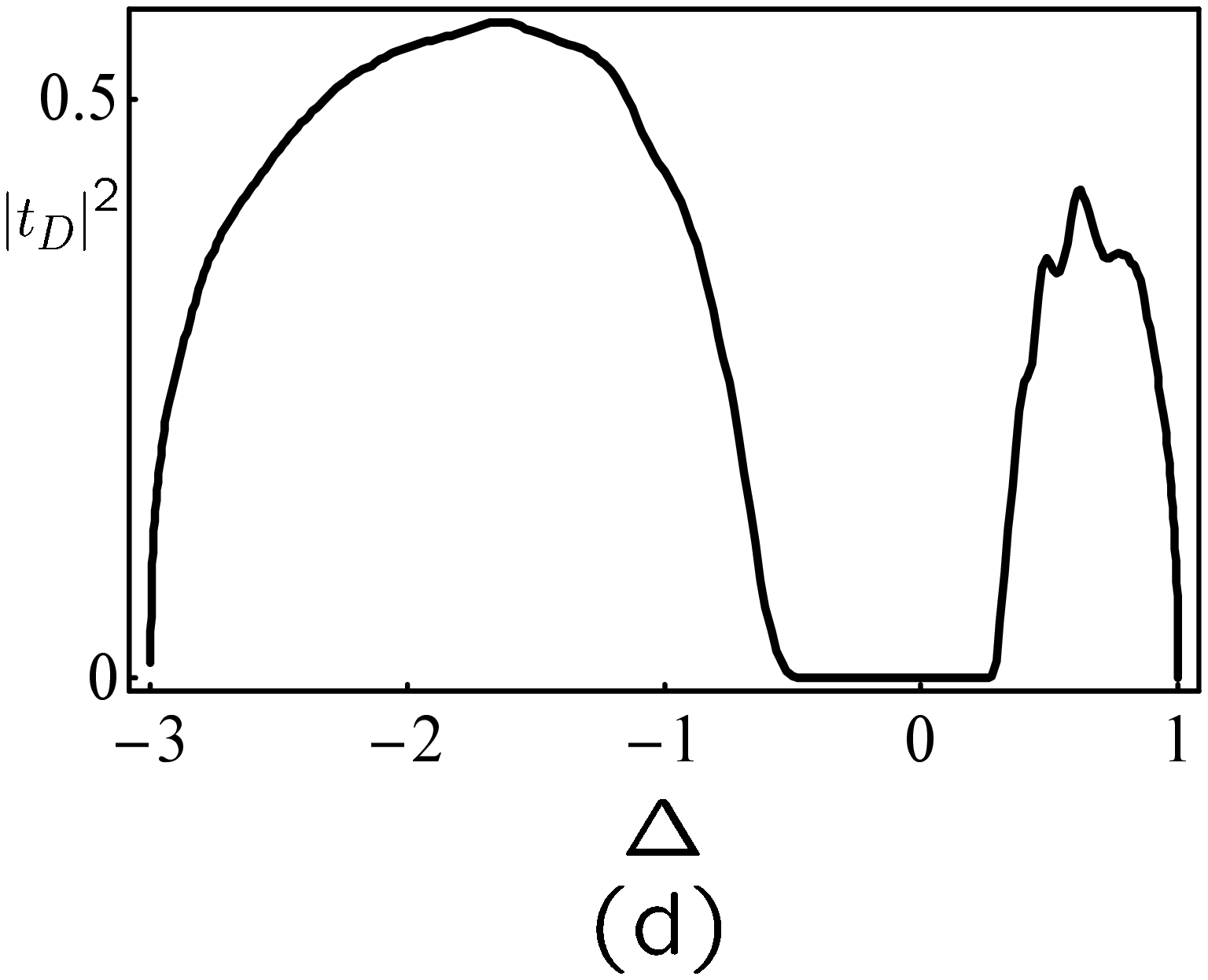}
\caption{The transmission coefficient $\left\vert t_{D}\right\vert ^{2}$ as
a function of the detuning $\Delta$ when the frequencies $\left\{ \protect%
\omega _{j}\right\}$ of the cavities and the coupling strength $\left\{
V_{ij}\right\} $ between the nearest two cavities are different from site to
site due to disorder. For given sets $\left\{ \protect\omega _{j}\right\}$
and $\left\{ V_{ij}\right\}$ [for (a)], the wide-band for perfect reflection
is almost unchanged while $\left\vert t_{D}\right\vert ^{2}$ changes
dramatically out of the wide-band. Figs. 9(b)-9(d) show the averaged results
of the $\left\{ \protect\omega _{j}\right\}$ and $\left\{ V_{ij}\right\}$
when the distributions of $\left\{ \protect\omega _{j}\right\}$ and $\left\{
V_{ij}\right\}$ are Gaussian with the same mean values $\protect\omega $ and
$V$, but different variances. When the variances of $\left\{ \protect\omega %
_{j}\right\}$ and $\left\{ V_{ij}\right\}$ are sufficiently small, e.g., $%
0.01\protect\omega$ and $0.01\left\vert V\right\vert $ [for (b)], the
transmission curve is quite similar to that in Fig. 3(c). When the variances
are increased, e.g., to $0.05\protect\omega $ and $0.05\left\vert
V\right\vert$ [for (c)], and $0.1\protect\omega$ and $0.1\left\vert
V\right\vert$ [for (d)], the range of the wide-band is shortened, and in
other region, the transmission curve is smoothed, with the depressed
envelop, which demonstrates the destruction of the coherence of the incident
photon due to the existence of the disorder.}
\end{figure}

The atomic decay and the loss of resonators also play essential role in
experiments. Usually, such atomic decay and resonator losses mean inelastic
scattering which results from the interaction between the system and a
realistic environment described by phonons. For simplicity, we investigate
this effect on the perfect reflection wide-band by phenomenologically adding
imaginary parts $-i\gamma _{a}$ and $-i\gamma _{c}$ to the two-level atom's
frequency $\Omega $ and coupled resonators' frequency $\omega $,
respectively. We note that these losses will can be directly added in the
final results (see Eq. (\ref{9})) phenomenologically. As a result, without
disorders, the transmission coefficient in Eq. (\ref{9}) becomes%
\begin{equation}
\left\vert t_{L}\right\vert ^{2}=\left\vert \frac{4V^{2}\sin k\sin
k_{L}^{\prime }}{C(E_{L})^{2}e^{ik_{L}^{\prime }\left( N_{a}-1\right)
}-D(E_{L})^{2}e^{-ik_{L}^{\prime }\left( N_{a}-1\right) }}\right\vert ^{2},
\end{equation}%
where%
\begin{equation}
C(E_{L})=Ve^{-ik_{L}^{\prime }}-Ve^{-ik}+w_{L}\left( E_{L}\right) ,
\end{equation}%
\begin{equation}
D(E_{L})=Ve^{ik_{L}^{\prime }}-Ve^{-ik}+w_{L}\left( E_{L}\right) ,
\end{equation}%
and $k_{L}^{\prime }$ satisfies the equation%
\begin{equation}
2V\cos k_{L}^{\prime }=2V\cos k-w_{L}\left( E\right) .
\end{equation}%
Here, $w_{L}\left( E_{L}\right) =g^{2}/\left[ E_{L}-\left( \Omega -i\gamma
_{a}\right) \right] $ and $E_{L}=\omega -i\gamma _{c}+2V\cos k$. The
corresponding reflection coefficient is%
\begin{equation}
\left\vert r_{L}\right\vert ^{2}=\left\vert \mathrm{i}2V\sin k\frac{%
C(E_{L})-D(E_{L})e^{-i2k_{L}^{\prime }\left( N_{a}-1\right) }}{%
C(E_{L})^{2}-D(E_{L})^{2}e^{-i2k_{L}^{\prime }\left( N_{a}-1\right) }}%
-1\right\vert ^{2},
\end{equation}%
With the same parameters in Fig. 3(c), we plot the transmission spectrum $%
\left\vert t_{L}\right\vert ^{2}$ and $\left\vert r_{L}\right\vert
^{2}+\left\vert t_{L}\right\vert ^{2}$ in Fig. 10, where $\gamma _{a}=0.02g$
and $\gamma _{c}=0.01g$. It is shown in Fig. 10(a) that the wide-band for
perfect reflection also exists in the transmission spectrum, with almost the
same boundaries when $\gamma _{a}/\Omega $, $\gamma _{c}/\omega \ll 1$, and $%
\gamma _{a}>\gamma _{c}$ are satisfied, but the envelop of the transmission
curve is depressed globally. This phenomenon also appears in Fig. 10(b),
which exhibits that the photon current is not conserved any more, especially
near the two boundaries $\Delta _{\pm }.$ However, when the single photon
stops in the interaction region, the energy is almost conserved. From the
above discussion,, the wide-band for near-perfect reflection is almost not
influenced by the imperfections such as disorder, atomic decay and resonator
losses existing in the realistic experiments..

\begin{figure}[ptb]
\includegraphics[bb=33 447 555 788, width=7 cm, clip]{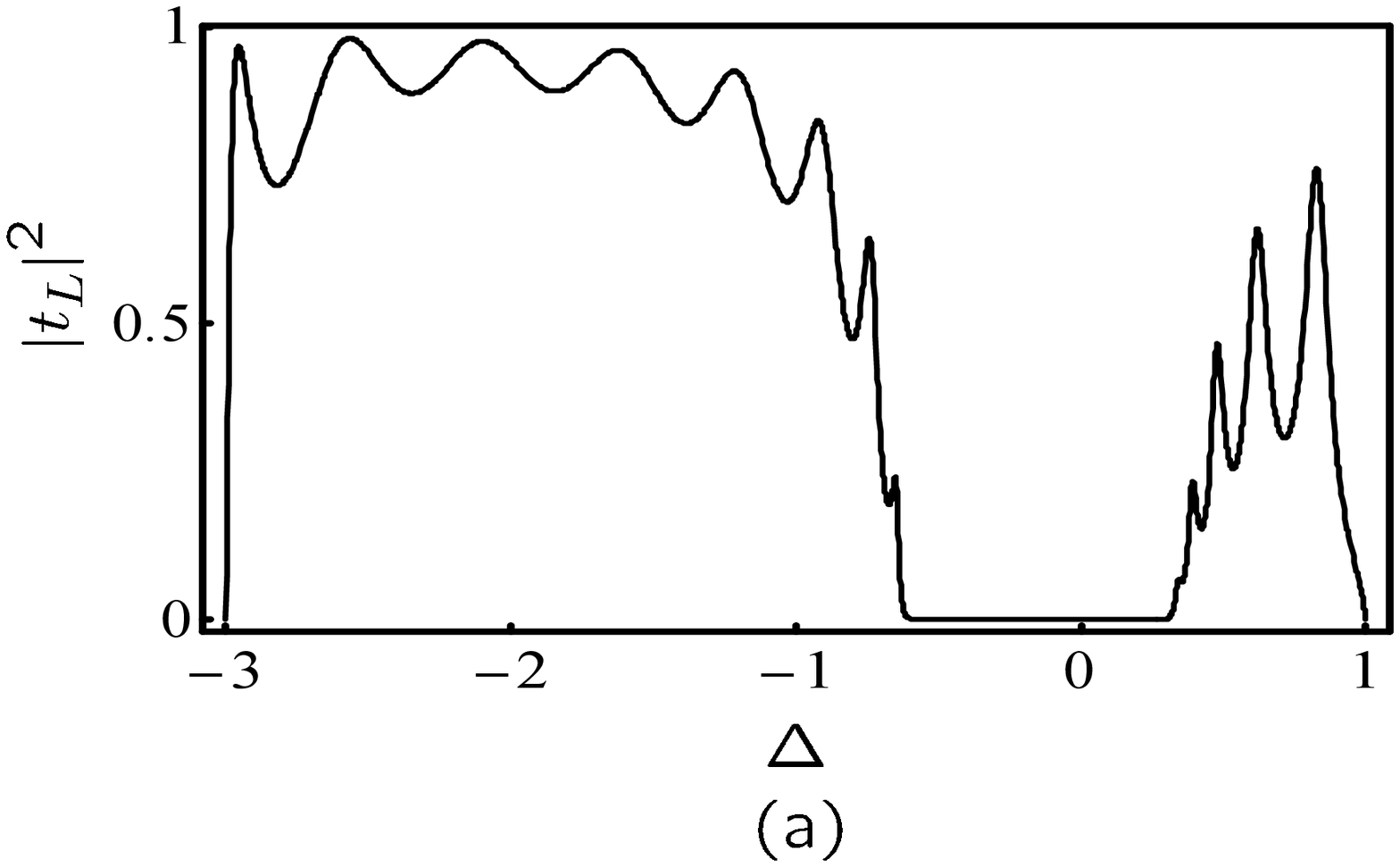} %
\includegraphics[bb=33 447 555 788, width=7 cm, clip]{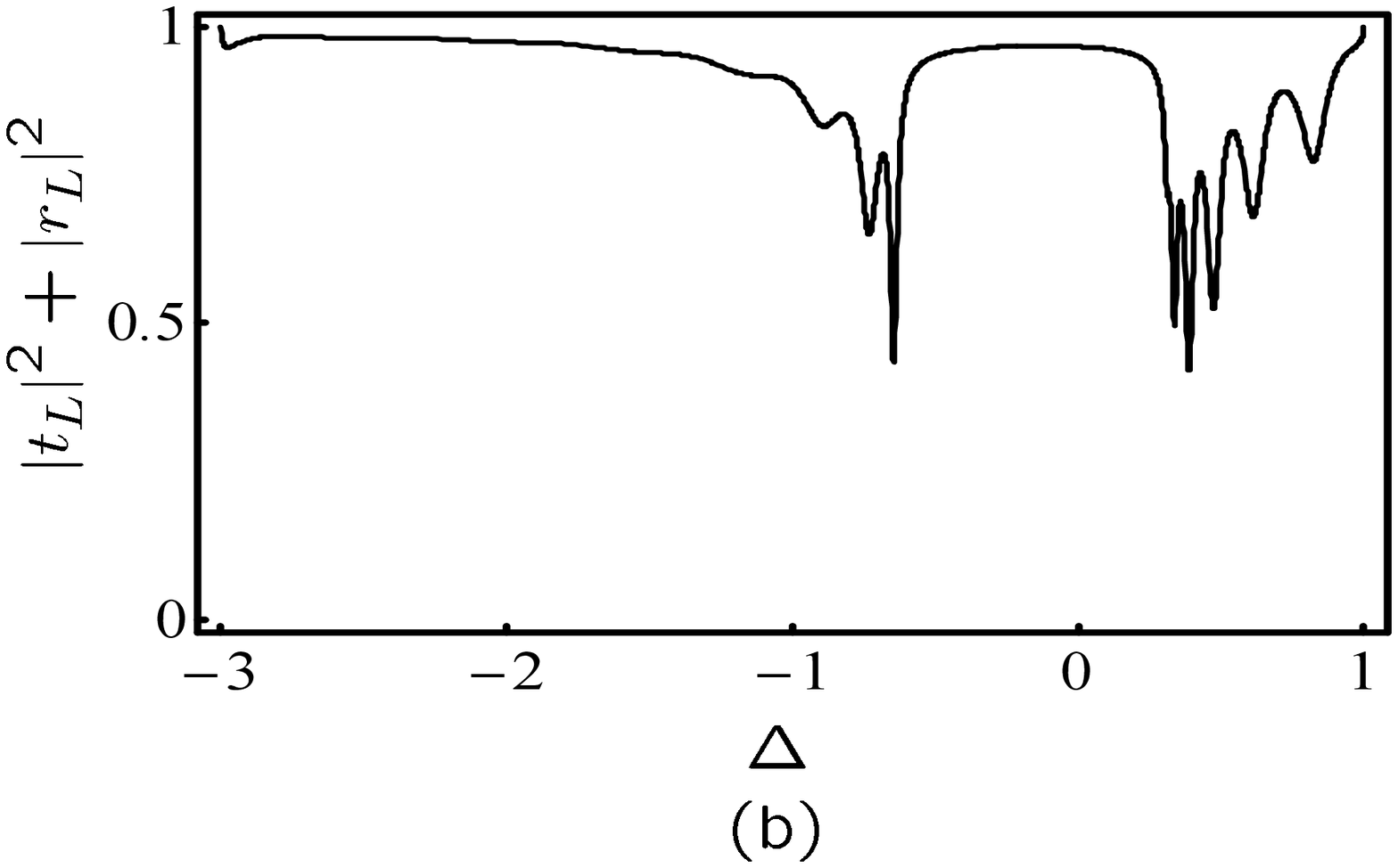}
\caption{The transmission coefficient $\left\vert t_{L}\right\vert
^{2}$ [for (a)] and $\left\vert r_{L}\right\vert ^{2}+\left\vert
t_{L}\right\vert ^{2}$ [for (b)] with respect to the detuning
$\Delta$ in the system-environment interacting case. It is shown
that the perfect reflection band still exist, and the photon current
is no more conserved.}
\end{figure}

\section{Summary}

We have studied the coherent transport of a single photon in a one-dimension
array \ of coupled-resonators individually coupled to two-level atoms. The
discrete coordinate scattering approach shows that a wide-band spectrum
appears for perfect reflection when the number of atoms $N_{a}$ is large.
The physical mechanism for this wide-band is considered by the incohenet
multi-reflection for light by $N_{a}$ atoms, which extends the the perfect
reflection line to form a wide band. The slowing and stopping light
phenomena also appear due to the interaction with atoms. We also diagonalize
exactly the photon-atoms interaction Hamiltonian in the interaction region,
and obtain two energy bands. It is found that the perfect reflection
wide-band is embedded in this band gap. We also consider the effect of
imperfections in experimental implements and find that the wide-band for
near-perfect reflection is not influenced when the parameters describing the
imperfections are small.

The model we proposed here can be realized by a circuit QED system \cite%
{wall,you1,you2,chio}, where the CRW can by realized by either defect
resonators in photonic crystals \cite{green} or coupled superconducting
transmission line resonators \cite{zhou3,liao,liao1}. By engineering the
photon-atoms coupling strengths and other parameters such as the hopping
constant and the energy space between the two levels of the atoms, we can
control the width and position of the perfect reflection wide band.

\acknowledgments The work is supported by NSFC No. 10474104, 60433050, and
No. 10704023, NFRPC No. 2006CB921205 and 2005CB724508.

\end{document}